# Sustainable Market Incentives: Lessons from European Feebates for a ZEV Future

Aditya Ramji[a b], Daniel Sperling[a], Lewis Fulton[a]

[a] *Institute of Transportation Studies, University of California, Suite #100, 1605 Tilia Street, Davis, California, USA – 95616 (dsperling@ucdavis.edu; lmfulton@ucdavis.edu)*

[b] *Corresponding Author: Aditya Ramji, adiramji@ucdavis.edu*

**Abstract**

Strong policies with sustainable incentives are needed to accelerate the EV transition. This paper assesses various feebate designs assessing recent policy evolution in five European countries. While there are key design elements that should be considered, there is no 'optimal' feebate design. Different policy objectives could be served by feebates influencing its design and effectiveness. Using feebates to transition to EVs has emerged a key objective. With the financial sustainability of EV incentive programs being questioned, a self-financing market mechanism could be the 'need of the hour' solution. Irrespective of the policy goals, a feebate will impact both the supply side, i.e., the automotive industry and the consumer side. Globally, feebates can be used to effect technology leapfrogging while navigating the political economy of clean transportation policy in different country contexts. This paper highlights thirteen design elements of an effective feebate policy that can serve as a foundation for policymakers.

## 1. Introduction

Global transport emissions will have to decline dramatically to achieve mid-century global net zero targets (International Energy Agency, 2021; UNFCCC, 2021a), and vehicle electrification is widely considered the most important strategy to achieve the goals (Ramji et al., 2021; "World Energy Outlook 2020 – Analysis - IEA," n.d.). Most national and global decarbonization strategies call for converting the vast majority of vehicles to electric propulsion by 2050, including virtually all light duty vehicle sales by 2040, or even earlier, in advanced economies (Brown et al., 2021; International Energy Agency, 2021; Miller et al., 2021; Sperling et al., 2020; UNFCCC, 2021b). While such ambitions exist, it requires a robust policy framework to facilitate and sustain this transition, with regulations aimed at a Zero Emission Vehicle (ZEV) transition as compared to the current focus on including plug-in hybrids as well (Cliff, n.d.).

Consumer purchase decisions are not purely a function of unit economics (Leard and McConnell, 2020), nor even TCO parity. Studies indicate the strong influence of broad attitudes about EVs in consumer choices (Choo and Mokhtarian, 2004; Egbue and Long, 2012; Lane et al., 2018; Orlov and Kallbekken, 2019), which are also influenced by many other factors including government pronouncements, regulatory requirements, and OEM marketing campaigns, all of which have an impact on the perceptions associated with EV purchase decisions (Larson et al., 2014; Xu et al., 2020; Ye et al., 2021; Zhang et al., 2018).

More recently, Germany announced the phase-out of certain EV incentives until funds last and have resulted in a decline in EV sales for September 2023 (EV-Volumes, n.d.; Reuters, 2022; Times of India, 2023). California, among the largest EV markets in the US and globally, has also decided to phase-out its flagship EV purchase incentives program, the Clean Vehicle Rebate Project (CVRP) in December 2022, citing fiscal limitations with the aim of reforming the incentives to focus only on middle- and low-income consumers (California Air Resources Board, 2023). China, the largest EV market globally, ended its purchase incentives program in 2022, and has now extended a 100% purchase tax exemption for EVs up to 2025, which will then be halved for 2026-27 (Li and Lee, 2023).

Many recent studies indicate that the timing of the (dis)incentive is critical in affecting vehicle purchase (including ZEV) choices, especially as upfront costs or rebates allow consumers to make firm decisions compared to delayed cash flows, which are then affected by capital availability and future cash flow discounting (Gong et al., 2020; Hardman et al., 2017; Jin et al., 2014; Yang et al., 2016). Imposing fees that impact the upfront vehicle purchase price have been found to have a greater impact on consumer choices (Li et al., 2019).

Studies have indicated the limitations of subsidies in achieving the intended outcomes of an EV transition (Edmunds, 2017; Gong et al., 2020), including issues of system efficiency and leakages, long-term sustainable market signals, and fiscal constraints (Thorne and Hughes, 2019; Wang et al., 2019). At the same time, the role of incentives cannot be undermined in facilitating the transition to electric vehicles (Lu et al., 2022) (Münzel et al., 2019). Even with enforceable ZEV mandates, the transition will require significant public finance support, and there are only a few policy options that governments can consider that are revenue neutral or surplus, and that can help finance the EV transition, including zoning fees, fuel taxes and feebates (Bose Styczynski and Hughes, 2019; Cansino et al., 2018).

The use of aggressive performance standards is widely considered a necessary condition to accelerate EV sales. Evidence from the world's largest EV market, China, also suggests the effectiveness of regulatory mechanisms over consumer subsidies in driving the shift to EVs (Deng and Tian, 2020; Li et al., 2020; Xiao et al., 2020). The EU has also used aggressive $CO_2$ vehicle performance standards as a centerpiece of their strategy to accelerate the transition (Dornoff et al., 2021; "EU vehicle targets | International Council on Clean Transportation," n.d.).

Governments have a variety of policy and regulatory instruments to draw upon to accelerate EV market sales, including subsidies, tax rebates, manufacturing incentives, and fuel efficiency standards. In practice, acceleration of sales will likely depend on a mix of these policies.

This paper makes a unique contribution to literature by way of a comprehensive review of feebate mechanisms in Europe, their evolution and potential impact in the post-2015 period of enhanced climate action and focus on ZEV transitions. In this paper, we consider the role of fiscal incentives combined with fees on performance standards, which are an important policy lever in generating demand and facilitating the ZEV transition. We analyze the evolution of the feebate schemes in different EU countries and identify key elements of a feebate design and its implementation that will be critical in ensuring that an accelerated ZEV transition is made feasible, in a cost-effective manner. It combines a unique vehicle make-model-powertrain analysis based on a combination of three different datasets that offer a more nuanced insight into feebate mechanisms in recent years, and their potential for replication in other geographies based on key policy objectives to be met. It also takes significance in the determining a framework for sustainable market incentives for electric vehicles that are not subject to annual budgetary fiscal constraints.

## 2. Feebates: advantages and design issues

A feebate mechanism essentially imposes a tax above a defined threshold value and offers rebates below the threshold. Feebates can be an effective tool to reinforce a level of carbon pricing across products and activities in multiple sectors including transportation, industry, electricity generation, electric appliances, and land use, among others (Batini et al., 2020; Scholz and Geissler, 2018). In this report, we focus on feebates for facilitating a ZEV transition. Feebates for vehicles are typically designed with two components: (i) a fee on the sale of vehicles that have higher rated $CO_2$ emissions (sometimes, may include additional metrics such as engine size or pollutant emissions as well) than an identified threshold level, and (ii) a rebate for the purchase of vehicles with emissions below this threshold. A pivot point or zero point is defined as the threshold above or below which fees or rebates would apply, with the fee estimated based on an efficiency or $CO_2$ criterion but also possibly adjusted by other vehicle attributes such as weight or footprint (German and Meszler, 2010).

A feebate can become an important policy tool, if its design is revenue-neutral, preserves consumer choice to the best extent possible, takes into consideration the industry's capabilities and is politically acceptable.

While feebates have been implemented in several countries and have been evaluated extensively in literature over the past decade, most of their focus has been on reduction of GHG and/or local pollutant emissions and inclusion of hybrids as a key technology transition choice while EV technology matures, rather than a focus on ZEV transitions.

A feebate policy instrument is compelling for a number of reasons: (i) it can be designed to be equitable and revenue neutral, with no burden on taxpayers (rather only a redistribution among consumers) and fiscal benefits for governments (*10*); (ii) it provides strong incentives to mitigate GHG emissions (Brand et al., 2013; Fazeli et al., 2017; Fridstrøm and Østli, 2017; Liu et al., 2012, 2011); (iii) it is a flexible and cost-effective regulation with continuity in funding streams, and is not subject to periodic appropriations decisions by legislatures and policy makers (which increases certainty for both consumers and manufacturers) (Kley et al., 2010); (iv) it harnesses market forces by adjusting price signals for vehicles (increasing the cost of ICE vehicles and reduces the cost of EVs); and thus, likely to gain support across much of the political spectrum for all the above reasons. Further, feebates, when complemented with state-level EV incentives, can accelerate EV adoption (Gillingham, 2013). Among other economic instruments, like purchase tax reduction, fuel tax, congestion charges, subsidies, and annual tax, feebates are emerging among the most promising policies to facilitate a ZEV transition, if designed efficiently (Antweiler and Gulati, 2013; Usher et al., 2015).

One of the limitations of feebates that is often argued is that benefits decrease with time as the increasing stringency of standards will lead to greater EV sales, resulting in lower revenue from fees (Liu et al., 2012), but as can be seen in discussions in subsequent sections of this paper, the French Bonus-Malus, , wherein the pivot point is revised periodically, provides a good case of balancing feebate cash flows and being revenue positive. At the same time, the objective of using feebates as a policy tool is to shape consumer behavior and achieve a critical mass of EV adoption, at which scale significant price benefits can be realized, thus, reducing the need for purchase incentives.

Studies (Berthold, 2019a; Kley et al., 2010) also find that feebate systems are sometimes seen as small-car subsidies, but it is important to understand that most early-stage EV models within affordable price ranges were smaller cars, given higher technology costs, a trend that is changing. While the transition to sustainable transportation will require consumers to shift to zero emission vehicles (ZEV), it does not necessarily mean a shift from an SUV to a small hatchback. Having said that, in the larger context of sustainable transportation and urban design, policies aimed towards a shift to smaller, more efficient cars may have larger co-benefits (Ellingsen et al., 2016; Higgins et al., 2017). Predicting consumer and industry reactions to outcomes such as ZEV sales, vehicle size and model availability, due to the applicability of feebates has been challenging leading to initial lag-periods in refining the feebate structure. This does call for better data to inform feebate policies.

Given the changes to feebate policies in recent years across countries, there is limited literature reviewing the design aspects of the revised mechanisms and the potential impacts towards a ZEV transition. A more recent 2019 study estimates that while the costs of a ZEV transition are significant, the benefits far outweigh the costs, as early as five years from implementing a $CO_2$-indexed fee structure for new vehicles (Slowik et al., 2019).

## 3. Data and Methodology

We compare the feebate mechanisms across five countries, namely, France, Sweden, Germany, Italy, and the United Kingdom (UK). For the automotive sales data, we use Marklines, IHS Markit, and EV Volumes. For vehicle emission data, we use the European Environment Agency database monitoring CO2 emissions from passenger cars. We also use the ACEA Tax Guide of multiple years to obtain data on vehicle taxation and EV incentives. For the respective feebate regulations, we obtain this information from the regulatory authority in each country that is administering the program.

For the analysis, we create a comprehensive and unique dataset of automotive LDV sales by make-model, powertrain (ICE, BEV, PHEV), and CO2 emission to create unique country profiles for 2010 - 2021. We then overlay this with the feebate design mechanisms to better understand the implementation of the policy and draw lessons in terms of best practices for an effective policy design that can help meet the objective of a transition to ZEVs.

## 4. Use of feebates in Europe

In the European region, a total of 23 countries (out of 31) have some form of emission-based taxation on either vehicle ownership or acquisition or both, such as Norway, Netherlands, and Spain, among many others (European Automobile Manufacturers' Association, 2021). There are four European countries, namely, France, Germany, Italy, and the United Kingdom that have experience with either a full or partial feebate mechanism and contribute to over 63% of the new vehicle sales in the European region (OICA Production Statistics, 2020, n.d.), and form the core of this review. More recently, Sweden (2.4% of new vehicle sales in the European region) has also adopted a pure feebate mechanism and has been included in this analysis.

**4.1 Review of the feebate mechanisms in selected countries**

As we review the feebate mechanisms of the select five countries (Table 1), we present a brief overview of the key elements of a feebate policy design. While the French Bonus-Malus was introduced in January 2008, and is considered a strong success, countries have revised their fee and rebate mechanisms since 2017, with some common elements emerging (Bose Styczynski and Hughes, 2019): (i) non-linear fee function, with a steep rise in fees for higher emission values; (ii) increase or maintaining rebates at 2020 levels to encourage EV adoption, with lesser fiscal constraints (revenue in-flow from fees); (iii) clear donut-hole in feebate structures; (iv) making PHEV performance parameters more stringent (such as higher Actual Electric Range or AER requirements) to be eligible for rebates, indicating a clear signal to push consumers to ZEVs.

A. Functional form and parameters: A feebate can have different functional forms depending on its objectives. They can be designed as step functions, wherein a fee or rebate is constant for a range of values, or they can be designed as continuous, non-linear functions with differing slopes at different threshold values. Feebates can also be designed as a hybrid wherein the fee and rebate curves follow different functional forms, which is the dominant practice, with rebates using step functions. In terms of the choice of parameters, a feebate with an emissions threshold in combination with specific criteria as vehicle weight, footprint, or engine displacement, remains technology agnostic to alternatives of diesel and gasoline powertrains (German and Meszler, 2010).

B. Pivot point and donut-hole: The pivot point decides who is taxed and who receives a rebate. It requires a good understanding of the nuances of consumer choices and market behavior in the automotive sector to determine the pivot point. In their paper, Greene et al. (D.L. Greene et al., 2005), find that the estimation of the pivot point and the slope of the curve for fees should account for consumers' valuation of fuel economy, which is typically only the first three years of savings, and not the entire life cycle of vehicle ownership.

**Table 1: Feebate mechanisms across key European countries in 2021** (European Automobile Manufacturers' Association, 2021)

| Country | Feebate type | Functional form and parameter | Pivot point | Fee structure | Rebate structure |
|---|---|---|---|---|---|
| France | Pure feebate | Continuous function for fee, step function for rebate;<br><br>$CO_2$ emissions ($gCO_2$/km) + Vehicle Weight (kgs) | 133 $gCO_2$/km | $CO_2$-based (non-linear curve) + vehicle weight (€10 per kg beyond 1800 kg; *w.e.f. 2022*) | € 6,000 for BEV with purchase price < € 45,000 & emissions < 20 $gCO_2$/km;<br><br>€ 3,000 additional for low-income households |
| Sweden | Pure feebate | Non-linear, piece-wise continuous function for fee, continuous rebate function;<br><br>$CO_2$ emissions ($gCO_2$/km) | 90 $gCO_2$/km | SEK 107 per $gCO_2$ if emission is between 90 – 130 $gCO_2$/km<br><br>SEK 132 per $gCO_2$ if emission > 130 gCO2/km | Graded rebates offered for all vehicles with emission < 90 $gCO_2$/km;<br><br>Maximum rebate of SEK 70,000, not exceeding 25% of the vehicle price |
| Germany ("BAFA - electromobility," n.d.; "Germany's vehicle tax system: Small steps towards future-proof incentives for low-emission vehicles \| International Council on Clean Transportation," n.d.) | Partial feebate | Non-linear, piece-wise continuous function for fee, step function for rebate;<br><br>Engine displacement + $CO_2$ emissions | 95 $gCO_2$/km | Tax on engine displacement +<br>$CO_2$ tax (€ 2 per $gCO_2$ > 95 $gCO_2$/km up to 116 $gCO_2$/km;<br><br>increases up to € 4 per $gCO_2$/km for emissions > 195 $gCO_2$/km) | Annual tax bonus of € 30 for emissions from 1 – 95 $gCO_2$/km;<br><br>€ 6,000 for BEVs and FCEVs, if purchase price < € 40,000, else, € 5,000;<br><br>In case of PHEVs, bonus will be €4,500 if purchase price < € 40,000, else, € 3,750 |
| Italy | Pure feebate | Step function for fee, discrete rebates<br><br>$CO_2$ emissions | 160 $gCO_2/km$ | $CO_2$ tax (step-wise) from € 1,100 to € 2,500 | € 8,000 for 0-20 $gCO_2/km$ with scrapping; € 4,500 for 21-60 $gCO_2/km$ with scrapping; € 2,000 for 61-135 $gCO_2/km$; purchase price < € 50,000 or <€ 40,000 if >61 $gCO_2/km$ |
| UK ("Low-emission vehicles eligible for a plug-in grant - GOV.UK," n.d.; "New or used car : Directgov - Find new cars and show fuel running costs," n.d.) | Partial feebate | Step function for fee, single rebate structure;<br><br>$CO_2$ emissions | 50 $gCO_2/km$ | GBP 10 for gasoline vehicles with emissions < 10 $gCO_2/km$; up to GBP 220 for vehicles emitting 150 $gCO_2/km$; up to GBP 1345 for vehicles emitting 200 $gCO_2/km$ | GBP 1,500 for vehicles with emissions < 50 $gCO_2/km$, and at least 70 miles of all-electric range; purchase price < GBP 35,000 |

### 4.2 Functional form and parameters

In France, the feebate mechanism was originally structured as a step function, with discrete amounts based on classification of vehicle emissions. Automotive manufacturers took advantage of the step functions by making marginal improvements in $CO_2$ emissions and making vehicles qualify for lesser fees or greater rebates. After multiple corrections of the step function, France made an important change to the bonus-malus scheme in 2017, wherein, the 'fee' line was converted to a continuous non-linear function (every marginal change in $CO_2$ emissions had a cost associated with it) while the 'rebate' line was maintained as a step function (Figure 1). Further, France introduced an additional fee, based on vehicle weight, with effect from 2022, where in, for every additional kilogram of weight over 1800 kg, a fee of €10 is being charged.

Italy also introduced a feebate mechanism in 2019 (Figure 3) structured as a step function (Asadollahi, 2021), similar to the French bonus-malus when it was first introduced in 2008, even though the disadvantages of a step function have been well established. While there was a significant jump in EV sales in 2020 and 2021 reaching 8.6% of new sales after the feebate was introduced (compared to less than 1% in 2019), it has remained flat around 8% in 2022 and 2023 (January – September) (EV-Volumes, n.d.). It is yet to be seen how the feebate will incentivize adoption of ZEVs in Italy, especially given that the rebates also apply to ICE vehicles with emissions up to 135 $gCO_2$/km, if they are being scrapped to purchase an EV.

Germany, on the other hand, does not have a pure feebate mechanism. It has introduced a separate bonus scheme to incentivize EV sales and has amended the motor vehicle tax code independently to introduce a more stringent $CO_2$-based taxation system. The German fee function is essentially non-linear and piece-wise continuous, with a graded fee per $gCO_2$. From 2020, Germany changed the "Umweltbonus" policy increasing the subsidy for EVs from the 2016 policy. Starting 2021, Germany has also imposed a revised $CO_2$-based vehicle tax, along with the tax based on engine displacement. As of 2020, vehicle purchase taxes were €2 per 100 cm3 engine displacement, in case of gasoline, and €9.5 per 100 cm,[3] in case of diesel, and in either case, for emissions greater than 95 $gCO_2$/km, there was a fixed fee of €2 per $gCO_2$ ("Finally catching up: What powers the EV uptake in Germany? | International Council on Clean Transportation," n.d.). The linear $CO_2$-based emission fee has now been amended with effect from 2021 to a non-linear, more stringent $CO_2$-based emissions fee, that ranges from €2 to €4 per additional $gCO_2$ above 95 $gCO_2$ (Figure 3). The German rebate in 2021 for BEV and FCEV purchases was at €9,000 (wherein, €6,000 are from federal funds, balance from the manufacturer), while the rebate for PHEVs is €6,750 subject to an emissions rate below 50 $gCO_2$/km, and an all-electric range (AER) of at least 60 km (European Automobile Manufacturers' Association, 2021). Further, there is a nominal €30 tax rebate for vehicles with emissions from 50 $gCO_2$/km to 95 $gCO_2$/km, to incentivize PHEV adoption.

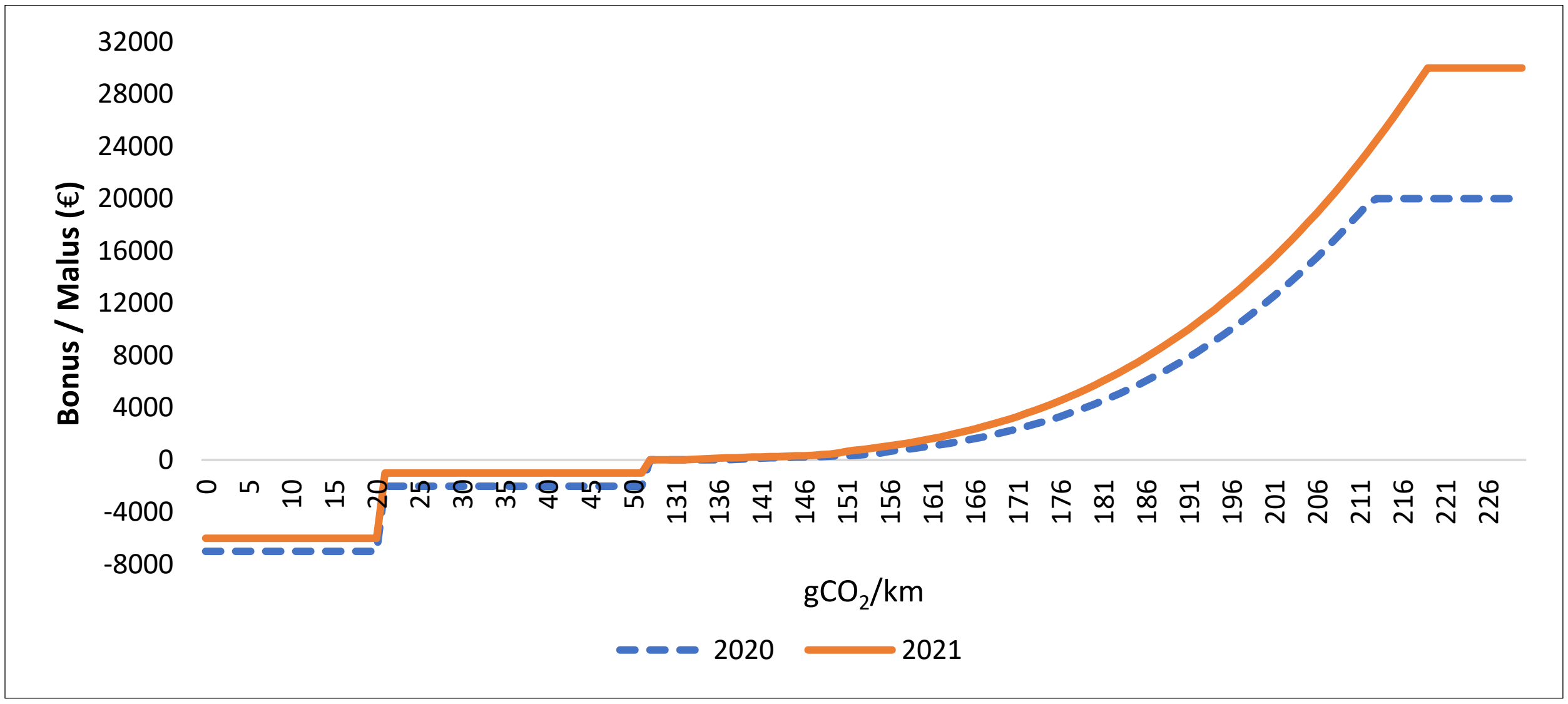

**Figure 1: Feebate functional form in France for 2020 and 2021**

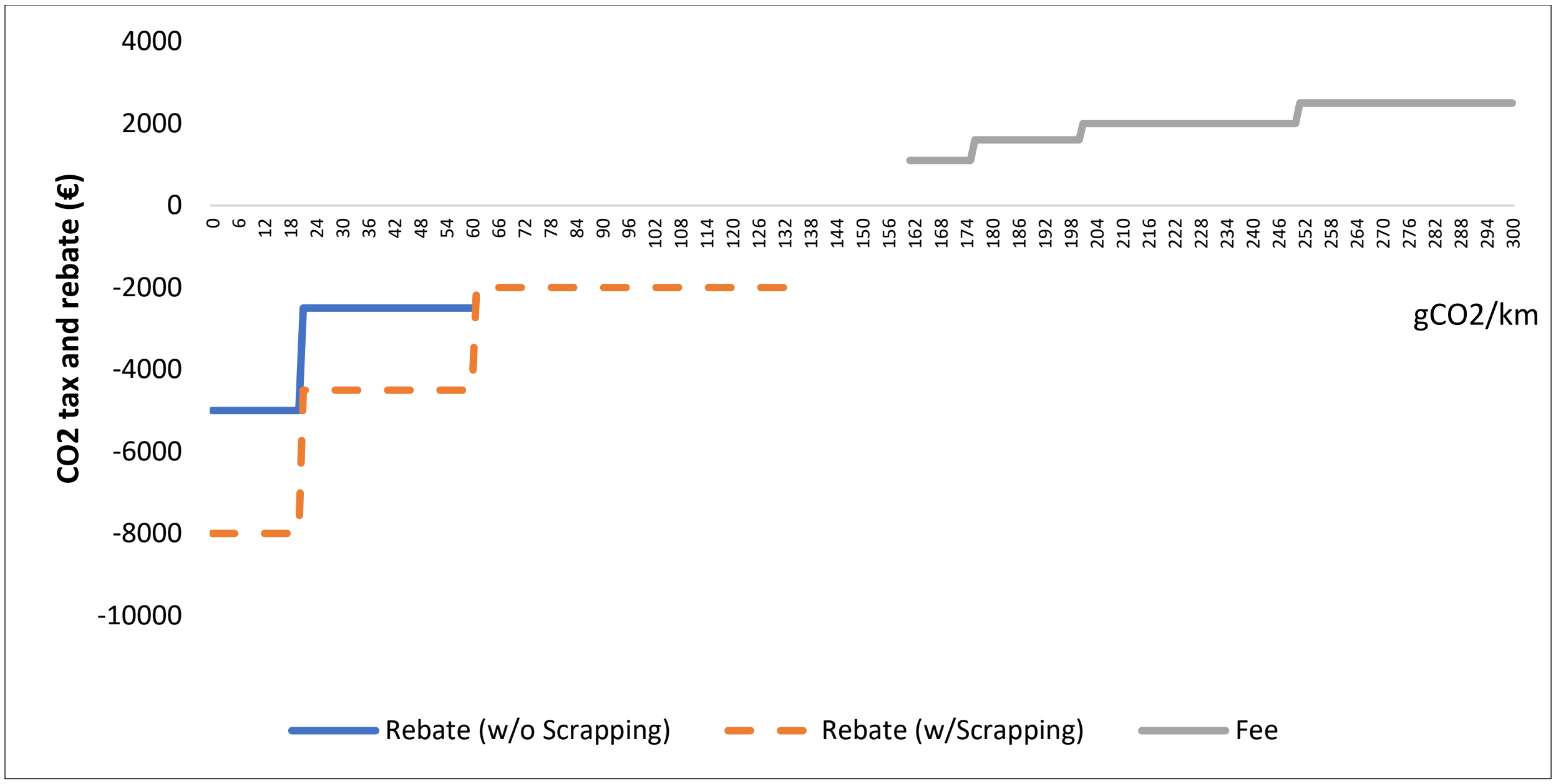

**Figure 2: Feebate functional form for Italy, 2021**

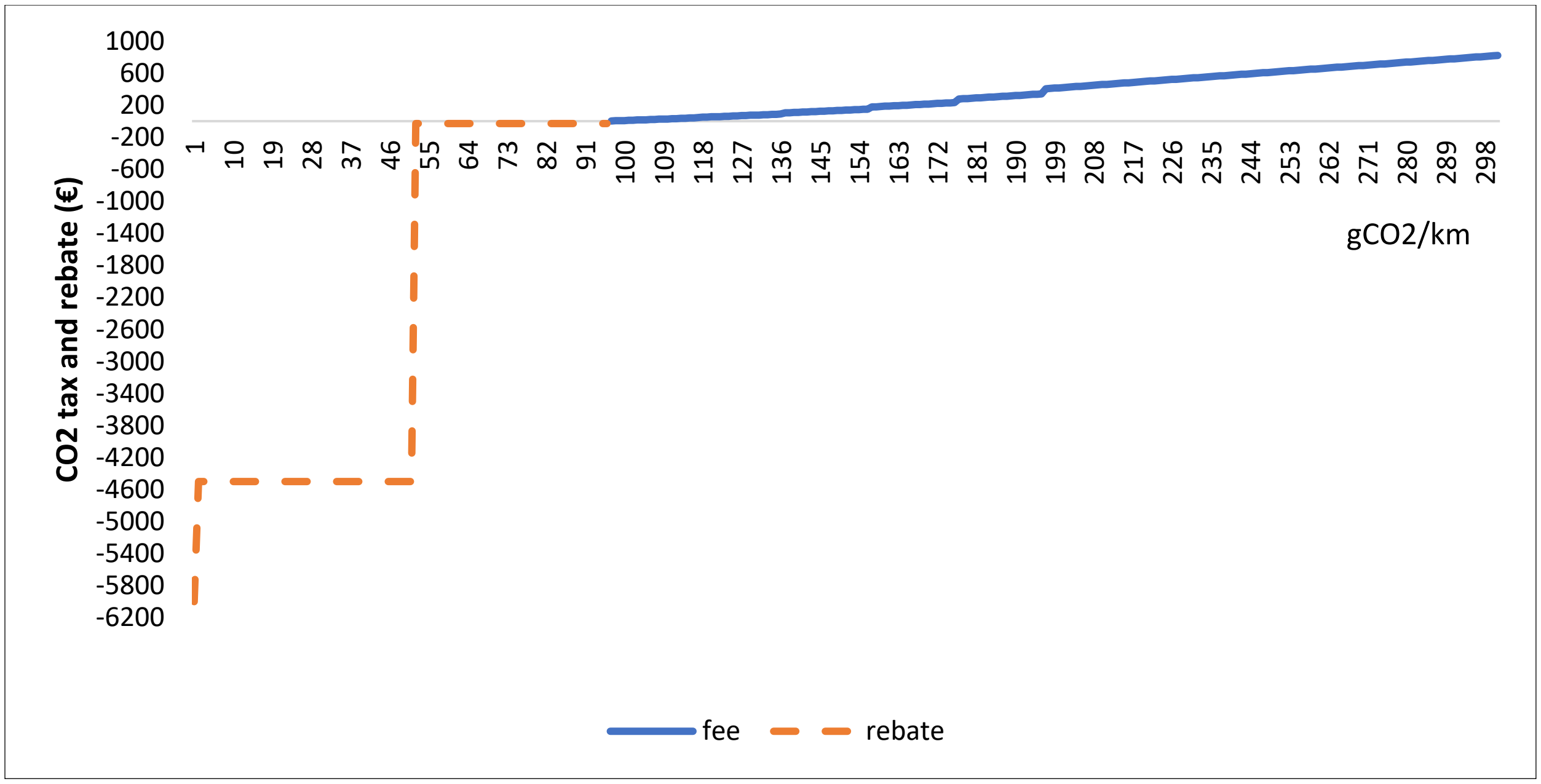

**Figure 3: Feebate functional form for Germany, 2021**

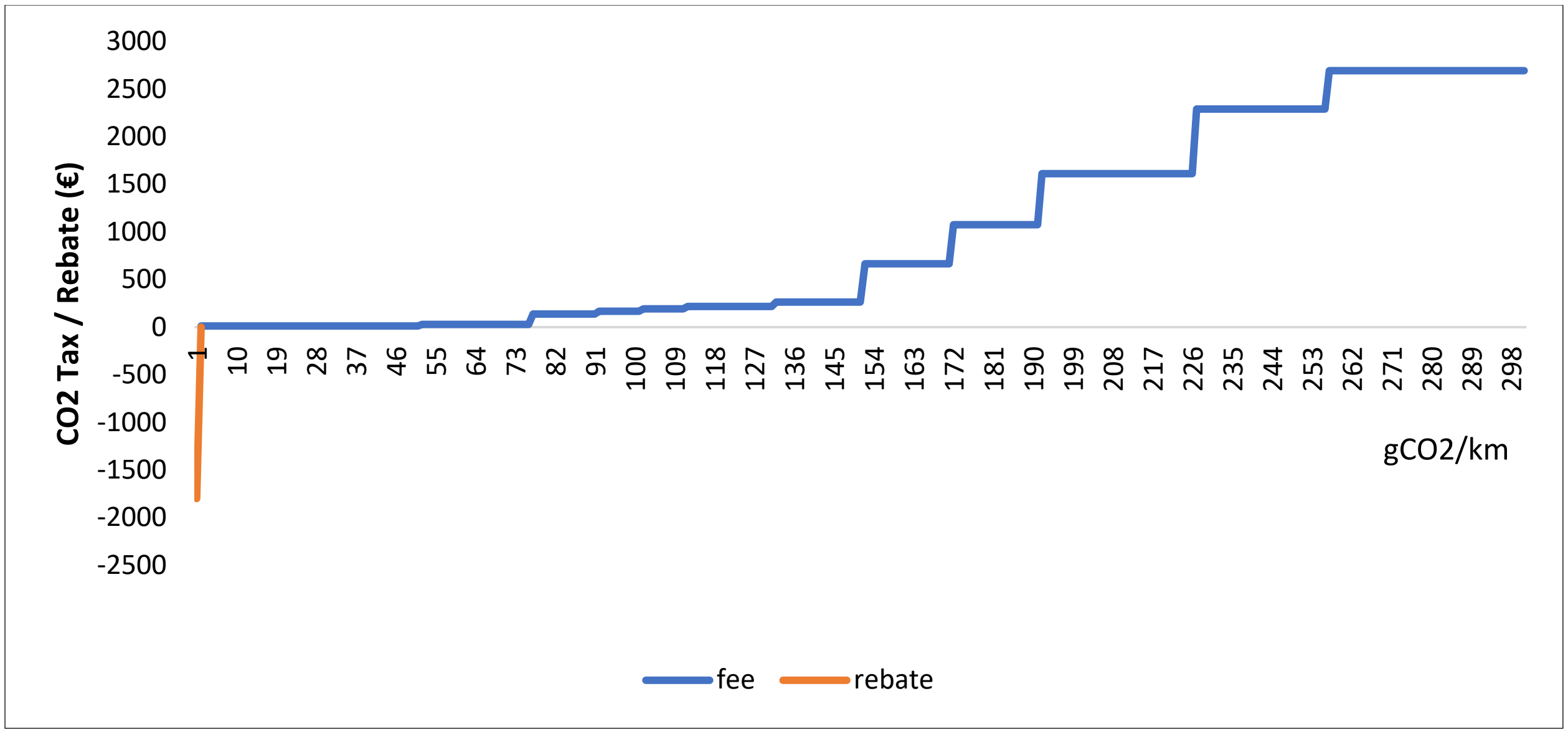

**Figure 4: Feebate functional form for the UK (1 GBP = 1.2 EUR), 2021**

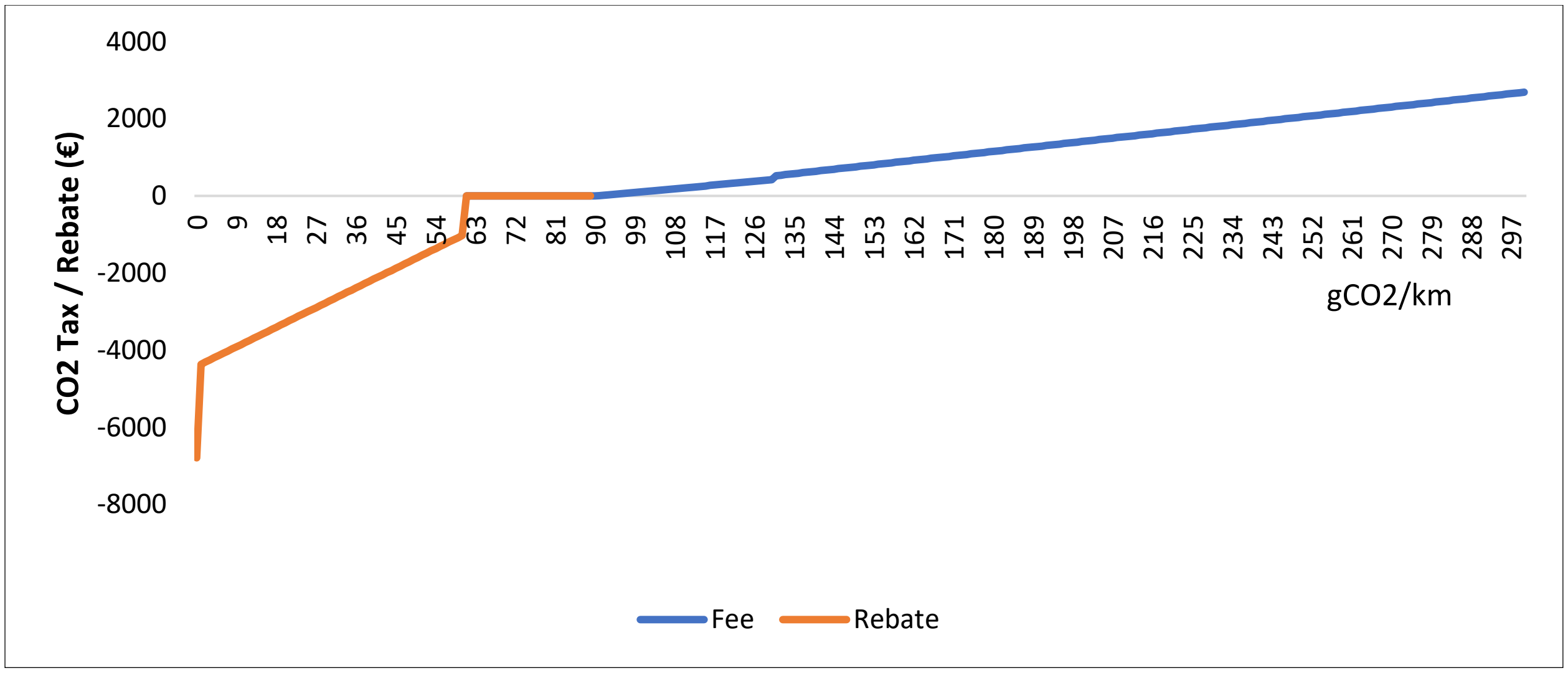

**Figure 5: Feebate functional form for Sweden (1 SEK = 0.097 EUR), 2021**

The UK also has a partial feebate mechanism with a $CO_2$-based vehicle taxation system, while the EV rebates are offered separately. The $CO_2$-based taxes are higher for diesel cars ("Vehicle tax rates (V149 and V149/1) - GOV.UK," n.d.). The UK emission fee follows a step function (Table 1 and Figure 4). The EV rebate on the other hand, has been reduced in 2021 from GBP 3,000 to GBP 1,500 for vehicles with emissions below 50 $gCO_2$/km, and with at least 70 miles of all-electric range ("Low-emission vehicles eligible for a plug-in grant - GOV.UK," n.d.). Further, the vehicle purchase price should not exceed GBP 35,000, with the grant not exceeding 35% of the purchase price. In addition, registration and ownership tax exemptions continue for zero-emission vehicles in the UK ("Country detail incentives | EAFO," n.d.).

In comparison, Sweden introduced the feebate mechanism on 1st July 2018, replacing a rebate-only program for green cars. The Swedish feebate follows a non-linear, piecewise continuous function for the emission fee, and is the only country among those being reviewed in this analysis to have a continuous rebate function (Figure 5). Prior to the feebate, ZEVs received a rebate of €3,800 while vehicles with emissions from 1 – 50 $gCO_2$/km, received a flat rebate of €1,800. Sweden's feebate structure (Table 1) is revised annually, with diesel vehicles liable to pay an additional fuel and environmental surcharge.

To provide a relative assessment of the impact the $CO_2$ fee has on purchase prices for consumers in the respective countries, Table 2 provides an insight with the VW Golf 2021 Gasoline model as a reference vehicle. We use the 2020 WLTP-based emission value for the vehicle in each country as provided in the European Environment Agency (EEA) database.

**Table 2: $CO_2$ fees for VW Golf 2021 across countries**

| *(€)* | **France** | **UK** | **Germany** | **Sweden** | **Italy** |
|---|---|---|---|---|---|
| MSRP* | 25445 | 31640 | 25445 | 34510 | 25445 |
| CO2 fee - Year 1 | 0 | 1074 | 70 | 91 | 0 |
| CO2 Fees - Year 2 to 4 | 0 | 0 | 210 | 216 | 0 |
| % fee on MSRP | 0.0% | 3.4% | 1.1% | 0.9% | 0.0% |
| gCO2/km (WLTP) | 119 | 171 | 127 | 102 | 121 |

**Maximum sales retail price; all monetary values are adjusted to Euros as per reference exchange rate*

**Table 3: Purchase rebate for Tesla Model 3 BEV and Ford Kuga PHEV 2021 across countries**

| | | **France (€)** | **UK (£)** | **Germany (€)** | **Sweden (SEK)** | **Italy (€)** |
|---|---|---|---|---|---|---|
| Tesla Model 3 BEV | MSRP (before rebate) | 43800 | 40490 | 39990 | 440000 | 35331 |
| | Rebate | 6000 | 1500 | 6000 | 70000 | 5000 |
| | MSRP (after rebate) | 37800 | 38990 | 33990 | 370000 | 30331 |
| | % rebate of MSRP | 14% | 4% | 15% | 16% | 14% |
| | *MSRP (after rebate) in €* | *37800* | *46285* | *33990* | *35980* | *30331* |
| Ford Kuga PHEV | MSRP (before rebate) | 40950 | 35915 | 39300 | 512700 | 36350 |
| | Rebate | 1000 | 0 | 4500 | 31552 | 2500 |
| | MSRP (after rebate) | 39950 | 35915 | 34800 | 481148 | 33850 |
| | % rebate of MSRP | 2% | 0% | 11% | 6% | 7% |
| | *MSRP (after rebate) in €* | *39950* | *42660* | *34800* | *46720* | *33850* |

As is evident from Table 2, the VW Golf 2021 model attracts no emission fee in France and Italy, as the vehicle emission value falls in the donut-hole of the feebate mechanism in both countries. The emission fee is the highest in the UK, followed by Germany and Sweden, as a share of the MSRP, which is driven essentially by the differential WLTP emission factor of the VW Golf in each country. More importantly, it should be noted that Germany and Sweden both impose the emission-based fee on an annual basis, compared to the other countries. As mentioned in the literature review earlier, the potential impact of a one-time initial higher fee compared to an annual fee on consumer choices is worth considering. As consumers tend to discount subsequent cash flows, an initial higher one-time emission fee can be the difference between consumers choosing a BEV or a PHEV or continuing with an ICE vehicle.

In Table 3, we can see that except for the UK, the BEV rebates are in the range of 14-16% across countries, while there is a wide variance with regards to rebates for PHEVs. In the case of PHEVs, we find that the rebate is highest in Germany at 11% of the MSRP, compared to about 6-7% in Sweden and Italy. It may be interpreted that France, Sweden and Italy are focused on incentivizing ZEVs over PHEVs, especially given that the top selling BEV and PHEV model compared are in similar price ranges.

In Table 4, based on the average WLTP emission of gasoline vehicles registered in 2020 in each country (as provided in the EEA database), we see the first-year emission fee, as per the current feebate mechanisms. With similar average vehicle emissions portfolio, the UK fee is significantly higher than that of Sweden and Germany. A more detailed analysis of how vehicle registrations are impacted by the emission fee is provided in the subsequent sections.

**Table 4: CO2 fees for year 1 based on average WLTP emissions of gasoline vehicles across countries**

| | France | UK | Germany | Sweden | Italy |
|---|---|---|---|---|---|
| Avg gCO2/km (WLTP) | 133 | 145 | 151 | 147 | 135 |
| CO2 fee - Year 1 | 50 | 264 | 140 | 71 | 0 |

In Table 5, we compare the five countries across key efficiency parameters that are currently being used or considered, to determine the feebate mechanism, i.e., $CO_2$ emissions, vehicle weight, engine capacity or displacement, and vehicle length. We use the NEDC test cycles for 2015 and 2020, for ease of comparison (with the WLTP test-cycle being completely applied from 2021).

**First, we can see average emissions from the NEDC cycle tests have improved considerably across all countries for all new cars sold between 2015 to 2020**, with Sweden showing a significant reduction at 6% CAGR, compared to an average of about 2% CAGR reduction in other countries.

**Second, with regards to vehicle mass, all countries have shown an increase in vehicle mass in the range of 1 – 1.5% CAGR on average**, with Sweden and the UK indicating the maximum increase, while Italy continues to have the lowest vehicle mass among the five countries. While France and Italy have relatively comparable vehicle parameters, the average emission reductions have been greater in France, and can probably be attributed to the longer running bonus-malus scheme, and more significantly, the higher malus component compared to Italy. In Figure 6, we plot the transition of countries from 2015 to 2020 across their average vehicle mass and NEDC emission ratings.

**Table 5: Country comparison across key efficiency parameters for feebate** (European Environment Agency, n.d.)

| *Country* | *Year* | **Avg NEDC ($gCO_2$/km)** | **Avg. Mass (kg)** | **Avg. Engine Capacity ($cm^3$)** | **Vehicle Size (mm)** |
|---|---|---|---|---|---|
| France | 2020 | 98.5 | 1360 | 1404 | 2613 |
| | 2015 | 111.0 | 1315 | 1481 | 2609 |
| Germany | 2020 | 113.6 | 1534 | 1698 | 2680 |
| | 2015 | 128.4 | 1447 | 1710 | 2643 |
| Italy | 2020 | 108.6 | 1351 | 1420 | 2573 |
| | 2015 | 115.6 | 1300 | 1464 | 2400 |
| Sweden | 2020 | 93.5 | 1656 | 1735 | 2733 |
| | 2015 | 126.3 | 1530 | 1773 | 2697 |
| UK | 2020 | 111.5 | 1510 | 1591 | 2678 |
| | 2015 | 121.3 | 1393 | 1635 | 2620 |

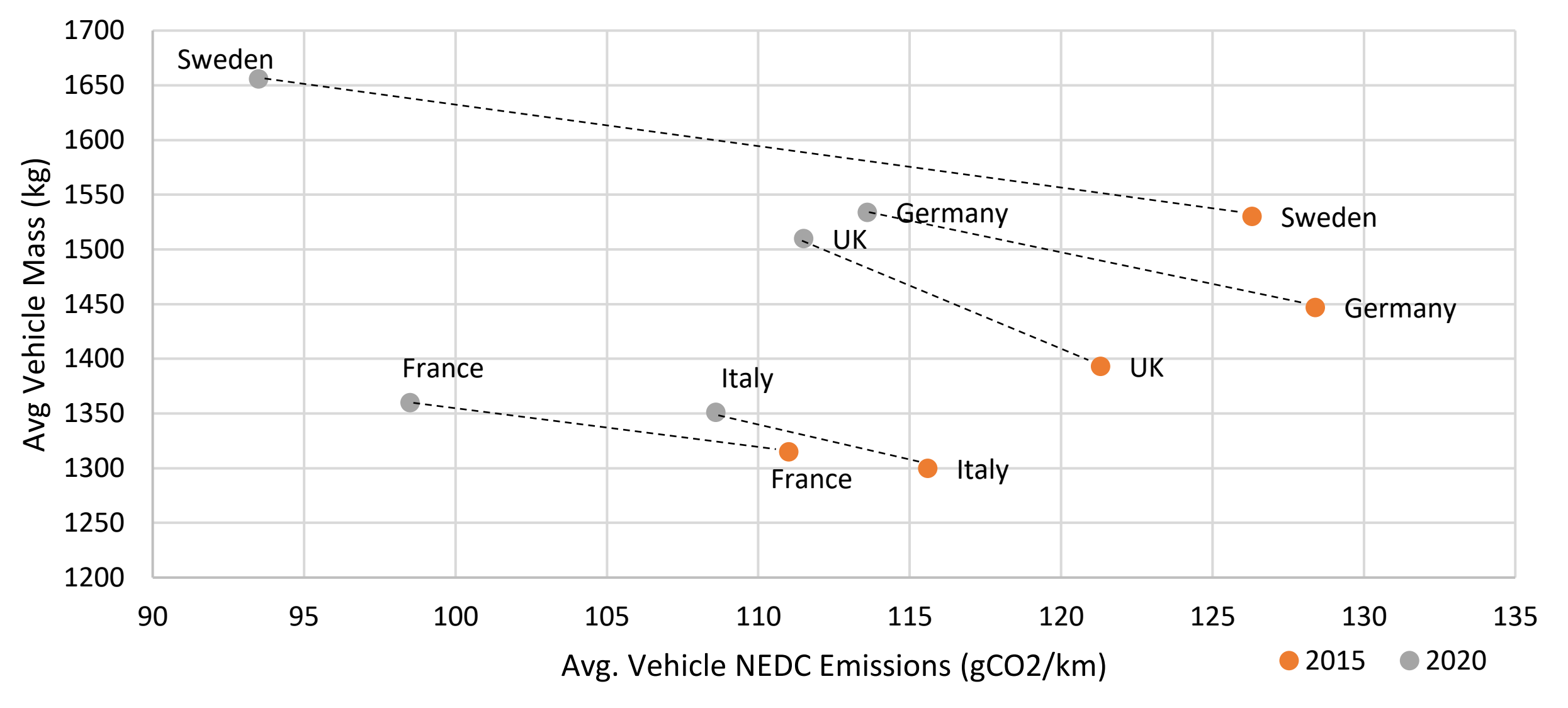

**Figure 6: Average vehicle mass and NEDC emissions across countries, 2015 - 2020**

With an average mass of 1360 kgs in 2020, the additional weight-based penalty introduced in France in 2022, will impact less than 5% of new vehicles sold, and is seen as a disincentive towards purchase of heavier and larger SUVs. Also, France has the highest average increase in vehicle mass (kg) to vehicle size (mm) ratio of 11.3 kgs/mm between 2015 to 2020, compared to an average of 2.6 kgs/mm among other countries in the same period, that possibly pushed French regulators to pre-empt automotive trends and introduce weight-based taxation measures.

**Third, it is interesting to note that average engine displacement has reduced across all countries, even though average vehicle mass and length have increased, between 2015 to 2020** (Table 5). Germany has an additional penalty on engine displacement for every $100cm^3$. It should be noted that engine displacement can broadly be seen as a measurement of engine power, and the power to weight ratio in an automobile is a good indicator of engine performance, speed, and acceleration. If we consider the engine displacement to vehicle mass ratio for 2015 and 2020 among the five countries, Germany remains the highest in both years. It is also interesting to note that Germany has been using engine displacement taxes since 2009 but has achieved the least reduction in power to weight ratio among the five countries between 2015 to 2020. But it should be noted that as the share of EVs increases, attributes of engine displacement could become redundant, and thus, there may be merits of having a single $CO_2$-based fee structure in the future. **Further, it is also seen that vehicle footprint in terms of average size has increased across all countries in varying magnitudes**, with France seeing only an average 4mm increase in vehicle size between 2015 and 2020, whereas Italy has seen the highest increase of 173mm, followed by the UK (58mm), Germany and Sweden (around 36mm) in the same period.

This raises the following key questions: (i) whether vehicle attribute-based taxation is effective or a simpler $CO_2$-based taxation with a large penalty amount is equally impactful or sufficient for a ZEV transition; and (ii) whether attribute-based taxation serves as a mechanism of hedonic pricing for emissions externalities where $CO_2$ taxation is difficult to implement.

**4.3 Choice of pivot point and donut-hole**

In the French Bonus-Malus scheme, the donut hole has been revised continuously. In 2008, the donut hole was between 125 $gCO_2/km$ - 160 $gCO_2/km$; in 2016, between 110 $gCO_2/km$ - 135 $gCO_2$/km. In 2021, the donut hole was 50 $gCO_2$/km – 133 $gCO_2$/km, with significantly more stringent vehicle penalties, capped at €30,000 above 219 gCO2/km. The rebate has also been decreased by €1,000 for 2021 compared to 2020. For the years 2022 and 2023, the upper threshold of the donut hole reduces to 128 $gCO_2$/km and 123 $gCO_2$/km, respectively, while the highest penalty cap will increase to €40,000 (>224 $gCO_2$/km) and € 50,000 (>226 $gCO_2$/km), respectively (European Automobile Manufacturers' Association, 2022).

In the Italian bonus-malus scheme, the donut hole currently stands between 136 $gCO_2$/km to 160 $gCO_2$/km. The fee is applicable as per a step-function for vehicles emitting above 160 $gCO_2$/km, with the highest fee at € 2,500 for all vehicles emitting more than 250 $gCO_2$/km. In the Swedish feebate mechanism, the "donut hole" is between 60 – 90 $gCO_2$/km, above which a fee applies, and below which a rebate applies.

In the German policy, although it is not a pure feebate, there was a donut hole up to 2021, wherein, there were no taxes for vehicles with emissions below 95 $gCO_2$/km, which has been replaced with a flat annual tax bonus of € 30 for emissions below 95 $gCO_2$/km. This can be considered a flat rebate "donut hole". In contrast, the UK taxation system has no donut hole as such, except that there is no taxation for hybrids that meet the criteria of emissions below 50 $gCO_2$/km and an all-electric range of at least 70 miles.

We now review the potential impacts of the feebate mechanism on vehicle registrations, emissions, consumer choices and their fiscal impact. The following discussion sections will evaluate the role of feebates in achieving a ZEV transition.

### 4.4 Impact on vehicle registrations and emissions

The choice of the functional form, efficiency parameter and the pivot point also bring into question which of the vehicles being sold are being taxed and which are benefitting from the rebate or being excluded from the feebate altogether. In France, while ICE vehicles were initially eligible to receive rebates, the scheme was revised in 2018 to make only EVs eligible for rebates, as is the case in all other countries.

We analyze the EV sales trend and the year of feebate changes for each of the countries (Figure 7). The year 2019 was an inflexion point across all five countries, with EV sales rising significantly in the following years, in 2020 and 2021. Also, it is observed that France, Germany and the UK follow a similar trajectory up to 2019, after which Germany has seen a sharp increase in EV market share in 2020-21 compared to France and the UK.

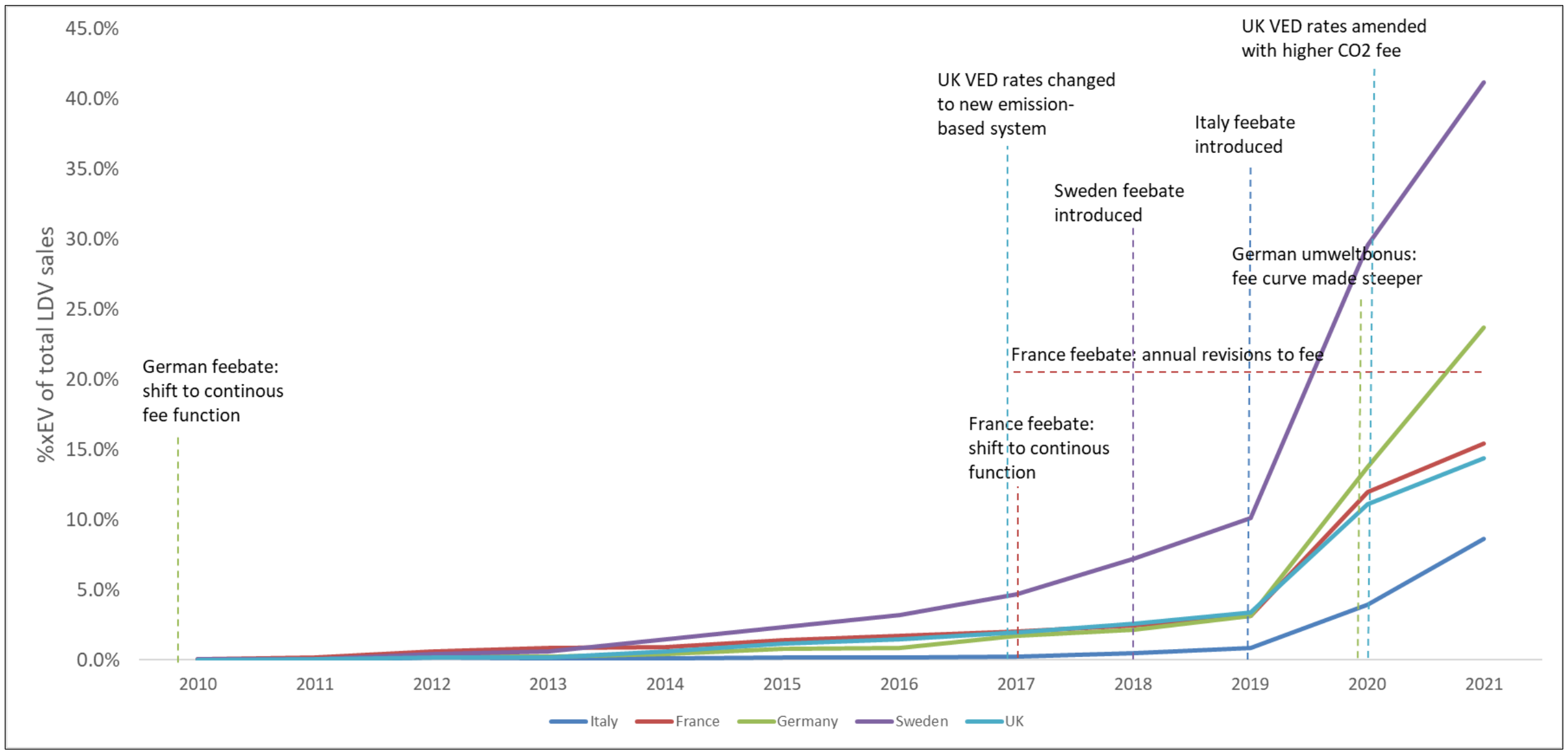

**Figure 7: Share of EV sales and changes in feebate mechanisms across countries, 2010-21**

While the EV share in total LDV sales reached around 15.4% in France and 14.4% in the UK for 2021, the share in Sweden jumped to ~41.2% in 2021, driven by a continued momentum in PHEV sales since 2020, and a significant increase in BEV sales in 2021. Italy has shown a doubling of market share in new EV sales for 2021, reaching 8.6%, compared to 4.3% in 2020, also driven by a greater increase in PHEV sales. In Germany, the EV market share reached 23.7% in 2021, compared to 13.7% in 2020. The role of EV incentives and the impact of the feebate mechanism in each of the countries provides important insights into the larger goal of achieving of a ZEV transition.

France and the UK are the only two countries with dominant BEV shares in 2020-2021, while PHEV sales have been increasing as well (Figures 8 and 9). In Germany, there has been a moderation in the growth of BEVs compared to PHEVs in total EV sales, although BEVs were marginally dominant in 2021 (Figure 10). There was an inflexion point in 2020, where the BEV and PHEV shares in total EV sales are equal in Germany. Italy is not very different, with equal shares of BEVs and PHEVs in total EV sales in 2021 (Figure 11). A recent study by ICCT indicates that the tax benefits introduced in 2019 for low and zero-emission company cars has boosted PHEV sales in Germany during 2019-20 (Bieker, 2019; "How to fix the plug-in hybrid loophole," n.d.; "More bang for the buck: A comparison of the life-cycle greenhouse gas emission benefits and incentives of plug-in hybrid and battery electric vehicles in Germany - International Council on Clean Transportation," n.d.). Sweden, on the other hand, remains the only country among the five, to have dominant PHEV sales since 2011, although BEV sales have grown steadily since 2018 (Figure 12).

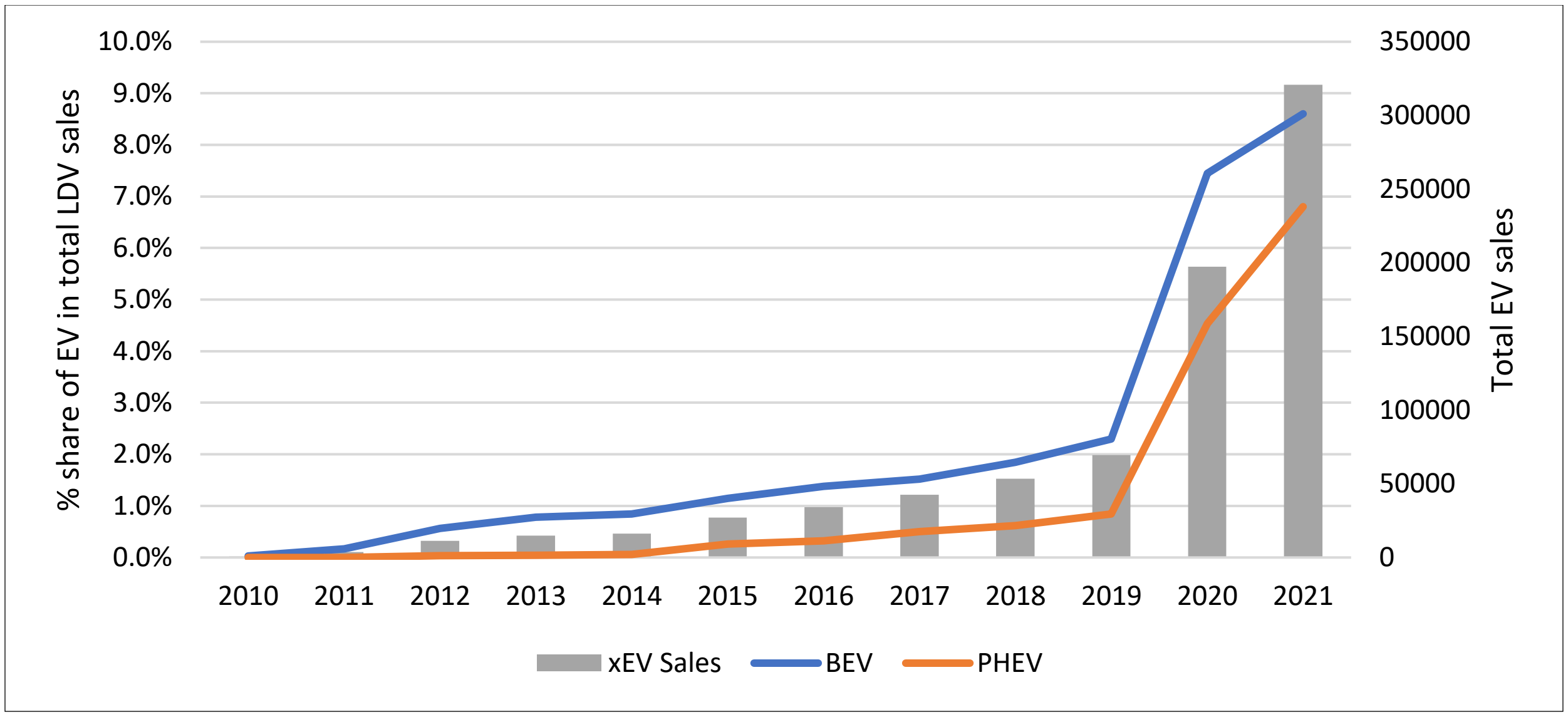

**Figure 8: Total EV sales and share of EV sales in total LDV Sales – France, 2010-21**

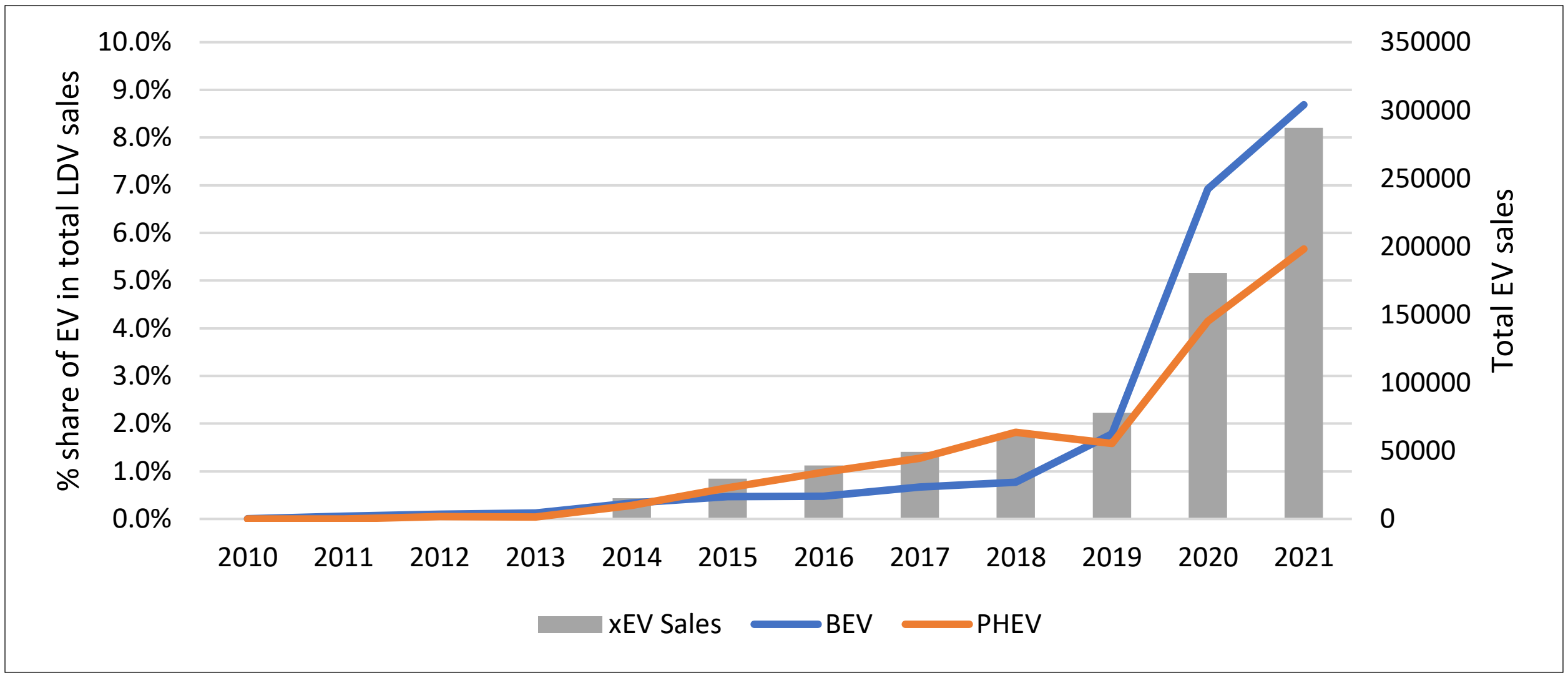

**Figure 9: Total EV sales and share of EV sales in total LDV Sales – United Kingdom (UK), 2010-21**

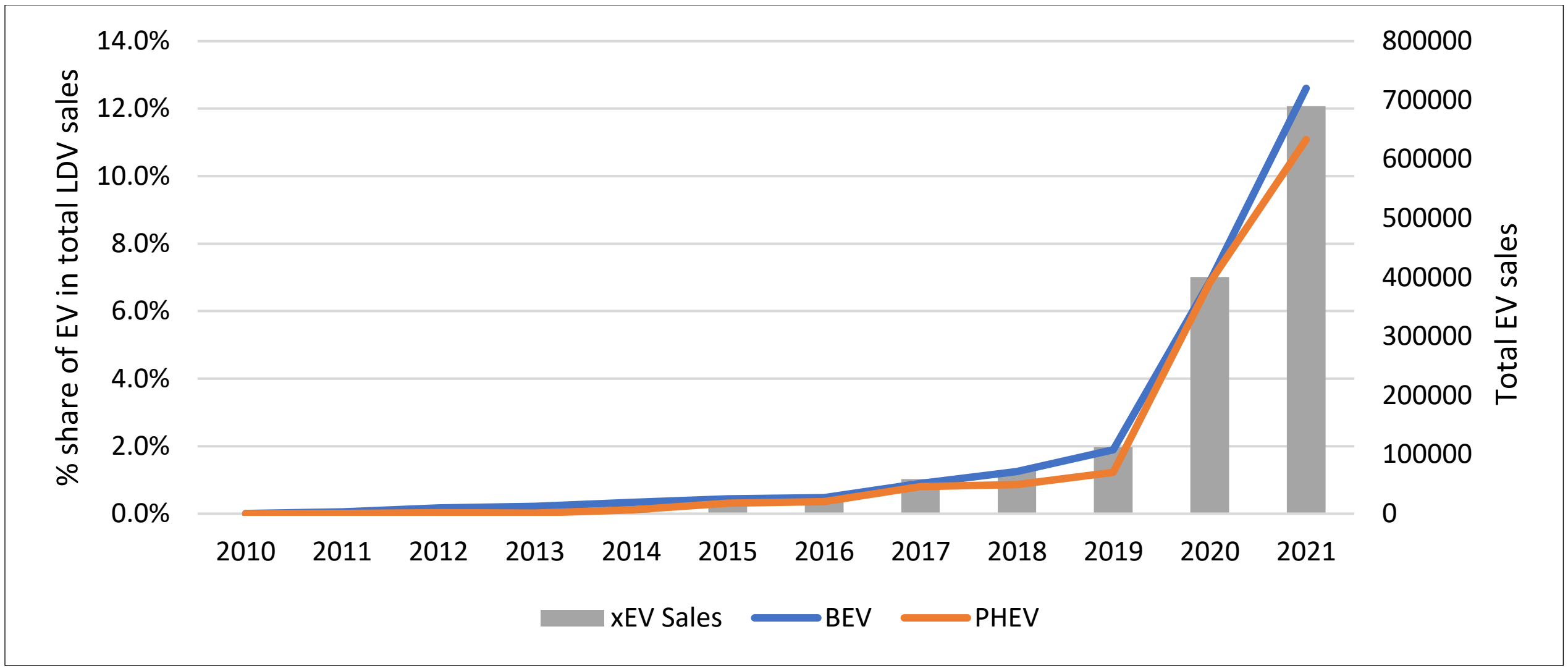

**Figure 10: Total EV sales and share of EV sales in total LDV Sales – Germany, 2010-21**

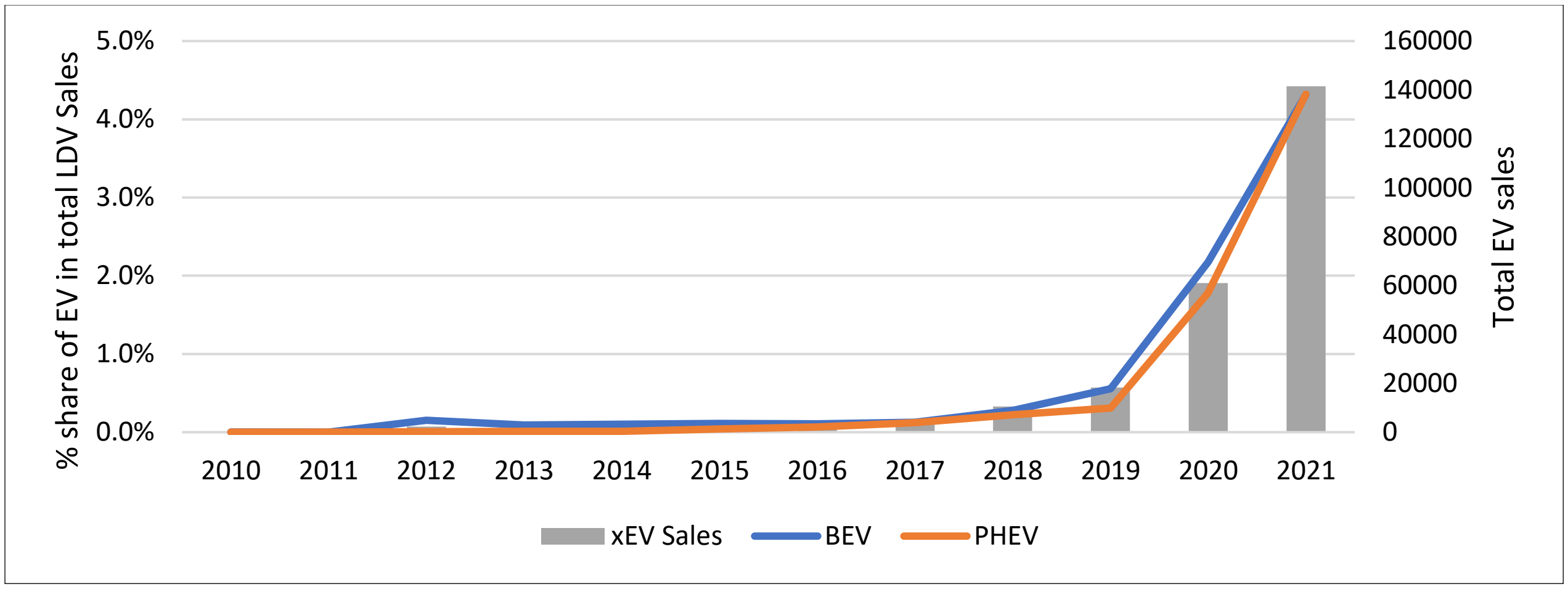

**Figure 11: Total EV sales and share of EV sales in total LDV Sales – Italy, 2010-21**

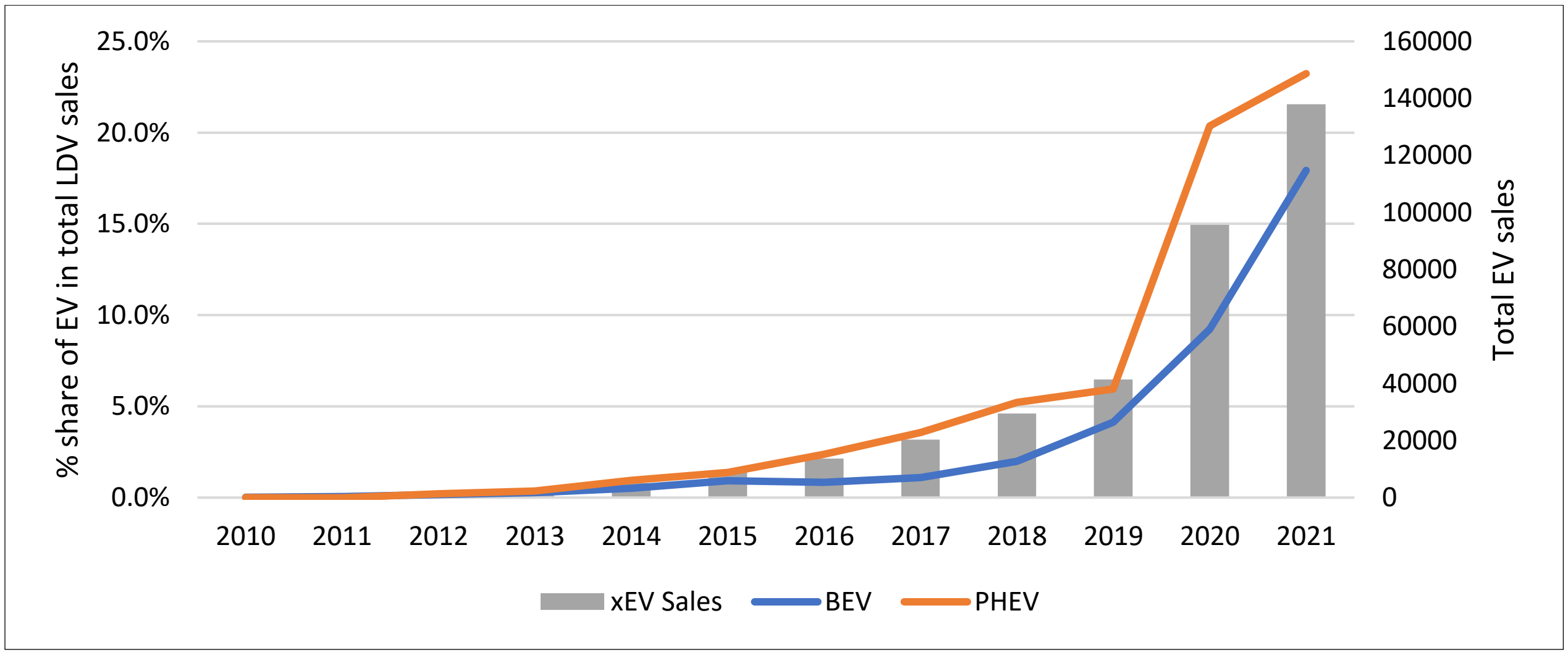

**Figure 12: Total EV sales and share of EV sales in total LDV Sales – Sweden, 2010-21**

Given the differences in EV sales across the five countries, we further investigate the model availability across BEVs and PHEVs, to see if model availability has posed any constraint to consumer choices.

As can be seen in Figure 13, between 2010 to 2017, the total number of EV models and variants sold across the five countries are relatively similar. Based on the data, we calculate the number of EV models in any given year if the volumes sold in that year is greater than zero (EV-Volumes, n.d.). It is only in 2018 that we begin to observe differences in model availability across countries. While the UK and Sweden lagged in terms of model availability between 2018-20 compared to the others, Sweden witnessed a significant increase in model availability in 2021 compared to the UK. Germany leads in terms of model availability since 2014 but takes a big leap in terms of availability from 2018-20, while France and Italy move similarly between 2017-20.

The availability of models alone does not drive higher shares of EV adoption and will depend on the other factors including vehicle prices, incentives, and disincentives, among others, including the feebate mechanism design in this case. For example, we observe that in 2021, Sweden has a lower availability of EV models and variants compared to most countries but has the highest market share (41.2%) of EV sales in total LDV sales for that year. The UK which has the lowest number of EV models sold in 2021, has an EV market share of total LDV sales of 14.4% in 2021, which is much higher than the 8.6% EV market share in Italy for 2021, even though the latter has a higher number of EV models sold for that year. From the earlier section on feebate reviews, we also know that the Italian feebate system has a much larger donut-hole than in the case of the UK, thus, impacting lesser vehicles within the scope of the emission tax, thereby impacting the rate of EV adoption.

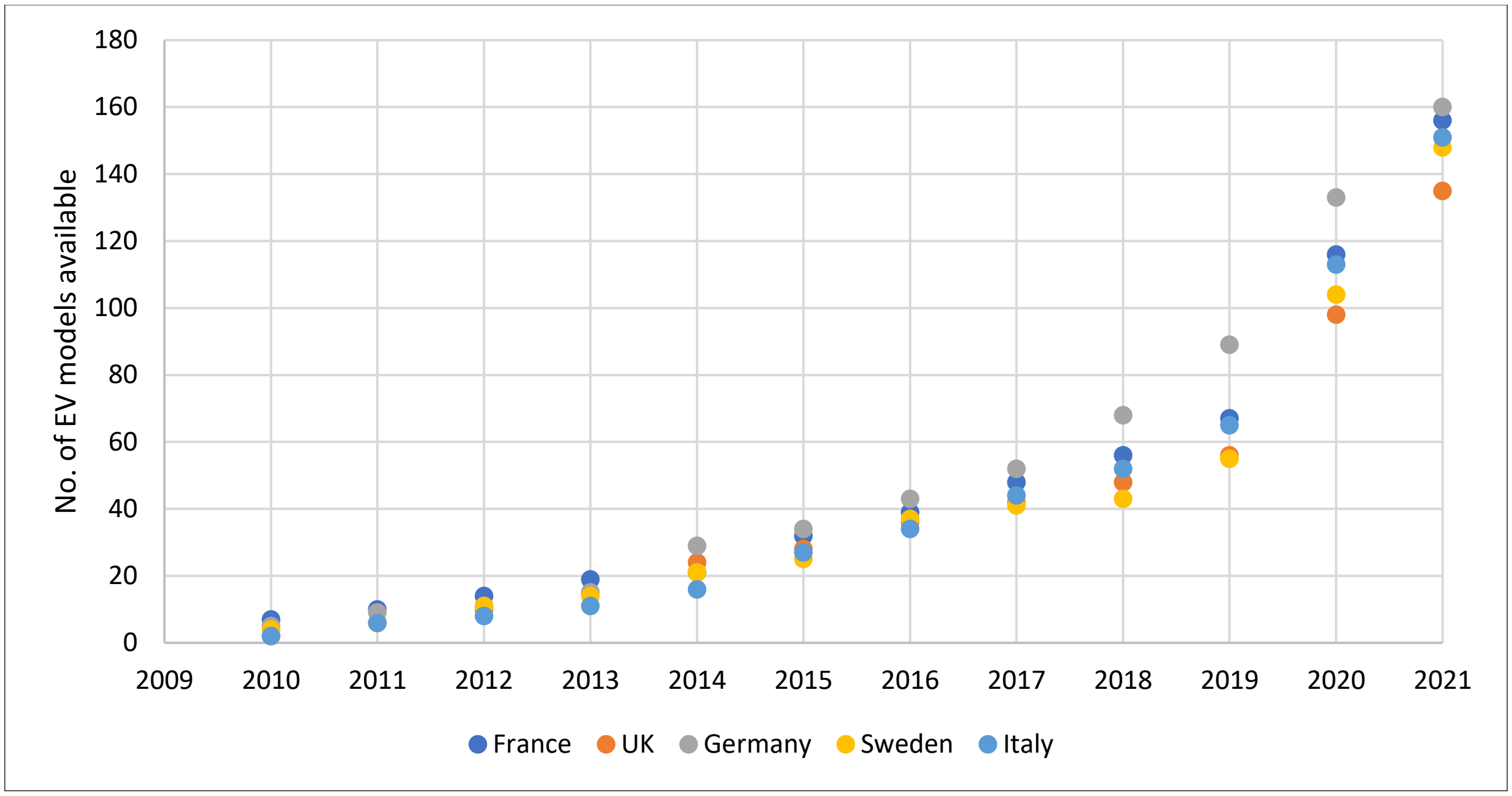

**Figure 13: EV model and variant sales by country, 2010 – 2021**

**Table 6: Country-wise availability of BEV and PHEV models between 2010 – 21**

| Country | EV | 2010 | 2011 | 2012 | 2013 | 2014 | 2015 | 2016 | 2017 | 2018 | 2019 | 2020 | 2021 |
|---|---|---|---|---|---|---|---|---|---|---|---|---|---|
| France | *BEV* | *7* | *10* | *10* | *13* | *14* | *16* | *17* | *17* | *21* | *27* | *48* | *66* |
| | PHEV | 0 | 0 | 4 | 6 | 7 | 15 | 20 | 29 | 32 | 38 | 66 | 87 |
| UK | *BEV* | *2* | *5* | *7* | *10* | *12* | *12* | *13* | *14* | *16* | *21* | *32* | *54* |
| | PHEV | 0 | 1 | 3 | 4 | 11 | 14 | 21 | 26 | 31 | 33 | 64 | 79 |
| Italy | *BEV* | *2* | *5* | *7* | *8* | *12* | *14* | *16* | *17* | *20* | *28* | *47* | *68* |
| | PHEV | 0 | 1 | 1 | 3 | 3 | 12 | 18 | 27 | 31 | 35 | 64 | 82 |
| Germany | *BEV* | *5* | *7* | *8* | *11* | *16* | *14* | *15* | *18* | *26* | *29* | *51* | *68* |
| | PHEV | 0 | 2 | 3 | 4 | 12 | 19 | 26 | 31 | 39 | 57 | 78 | 89 |
| Sweden | *BEV* | *4* | *4* | *6* | *8* | *13* | *9* | *13* | *11* | *15* | *20* | *37* | *60* |
| | PHEV | 0 | 2 | 5 | 6 | 8 | 15 | 22 | 28 | 27 | 34 | 66 | 86 |

**The average ratio of PHEV to BEV models available has been declining, indicating more BEV models being made available in the automotive markets across the five countries.** From an average ratio of 1.6 (PHEV to BEV models available in a year) between 2017-20, indicating the higher availability of PHEV models and variants, the ratio has declined to 1.35 in 2021, marking a rise in BEV model availability (Table 6).

A further analysis of the sensitivity of EV sales to model availability can be seen in Figures 14 – 18. We plot the EV sales to model availability for 2015 to 2021 and fit a linear trendline with an intercept to further understand the correlation (but not implying any causality). We observe that for every additional model available, BEV sales indicate higher growth propensity compared to PHEV sales among all countries except in Sweden. We observe that in Sweden the model availability to additional EV sales is relatively similar for both BEV and PHEVs, i.e., the slope of the trendlines for BEV and PHEV sales are almost coinciding. The ratio of incremental BEV sales per additional model available to PHEV sales per additional model available as shown in Table 7, clearly shows the response of EV sales to model availability. For every additional model available, the response ratio of BEV sales to PHEV sales is highest for the UK and France, followed by Germany and Italy.

**Table 7: Country-wise EV sales and model availability from 2015-21**

| **Country** | **BEV sales per additional model** | **PHEV sales per additional model** | **Sensitivity difference *(BEV per model / PHEV per model)*** |
|---|---|---|---|
| France | 2457 | 1191 | 2.1 |
| UK | 2860 | 1275 | 2.2 |
| Germany | 4138 | 2375 | 1.7 |
| Sweden | 889 | 871 | 1.0 |
| Italy | 792 | 566 | 1.4 |

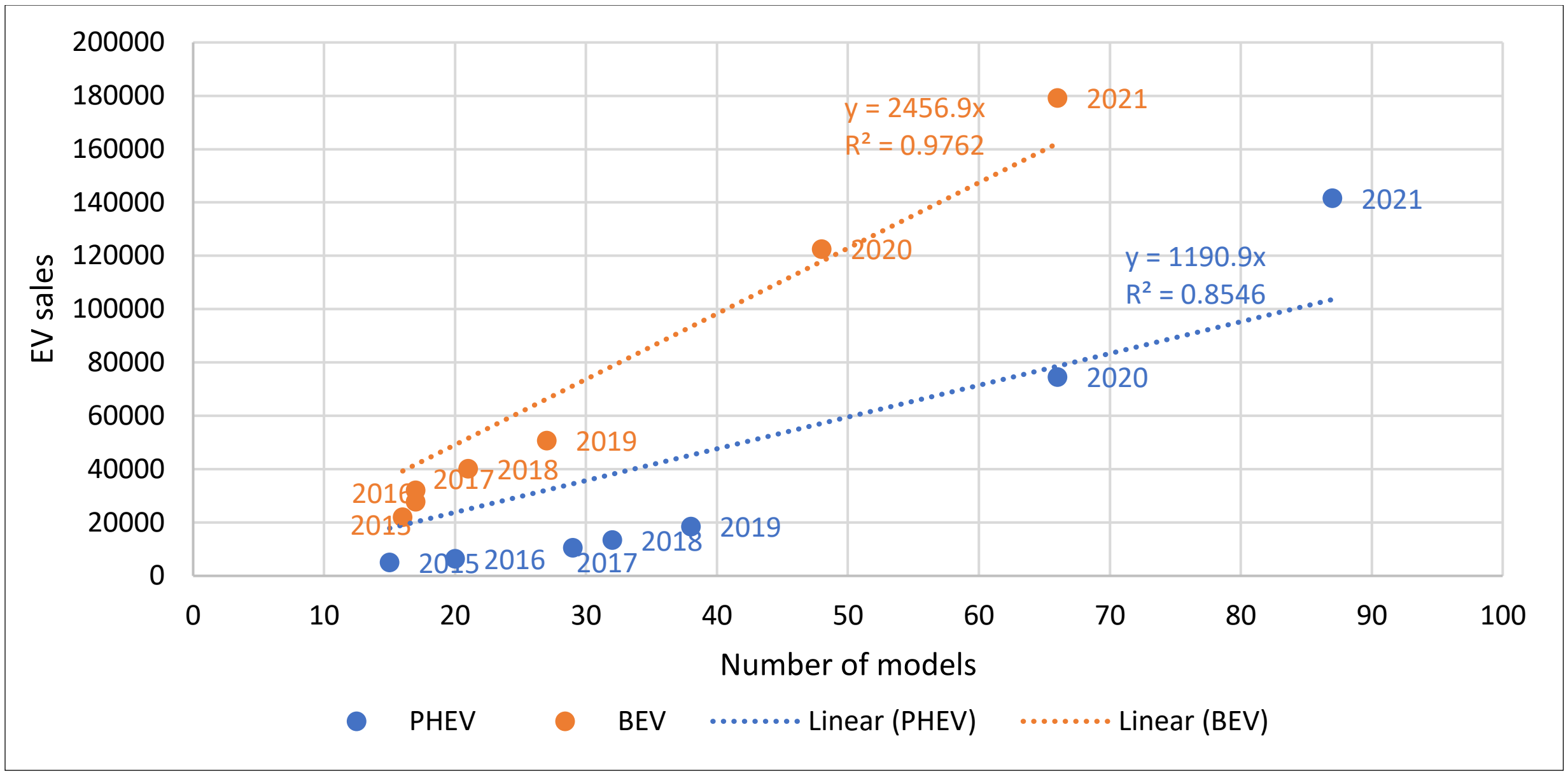

**Figure 14: EV sales vs model availability sensitivity in France, 2015-21**

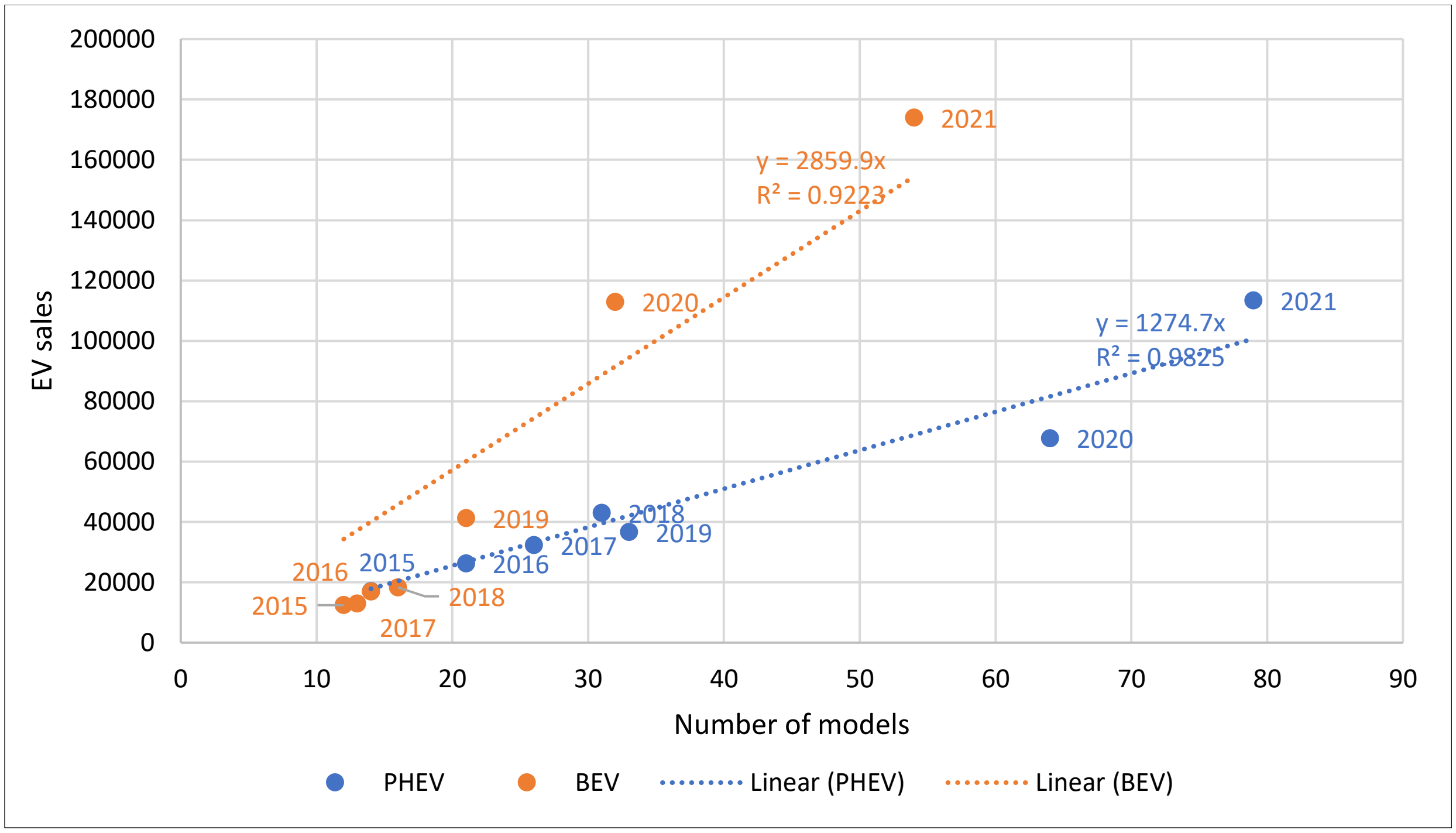

**Figure 15: EV sales vs model availability sensitivity in the UK, 2015-21**

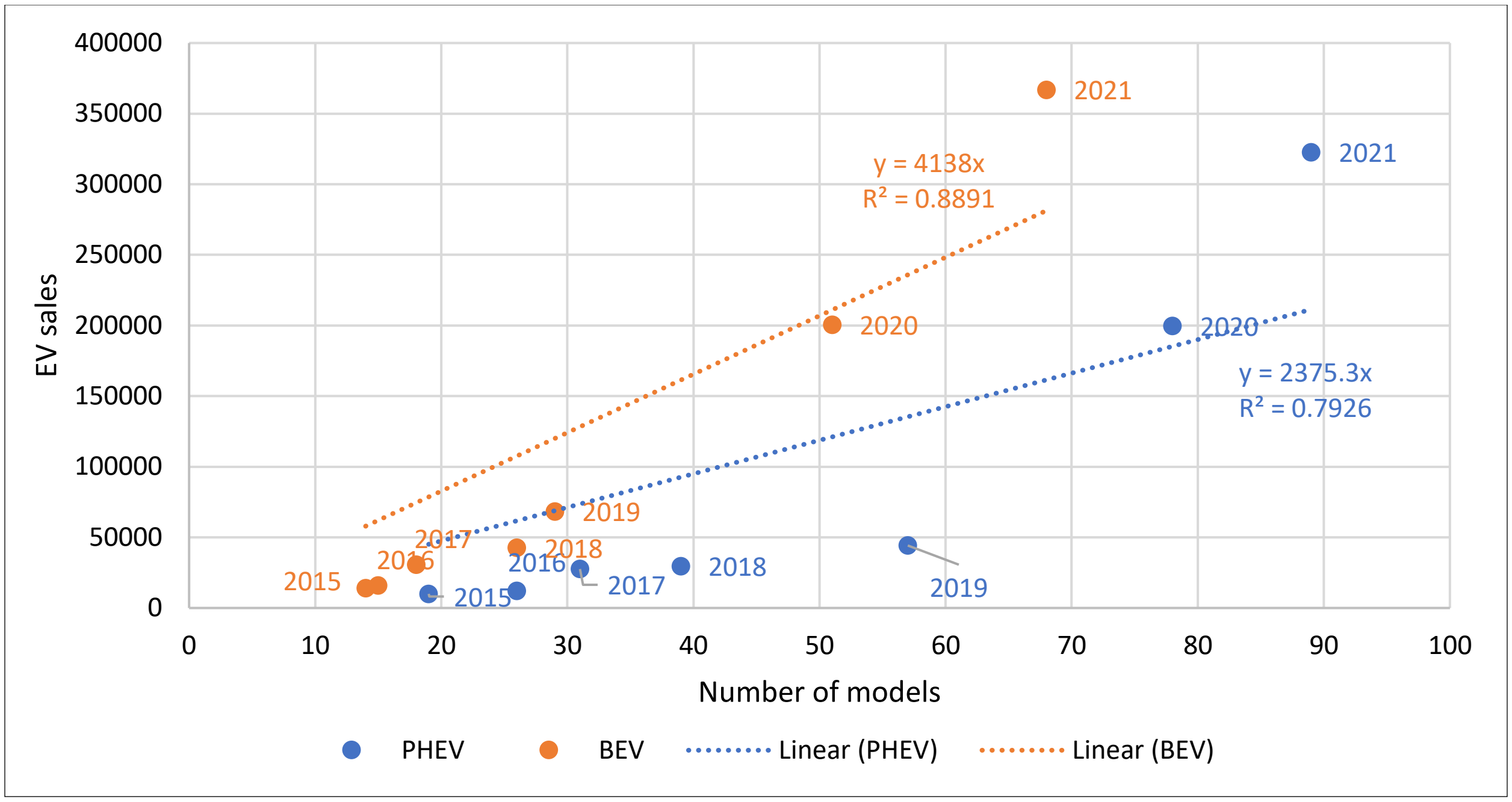

**Figure 16: EV sales vs model availability sensitivity in Germany, 2015-21**

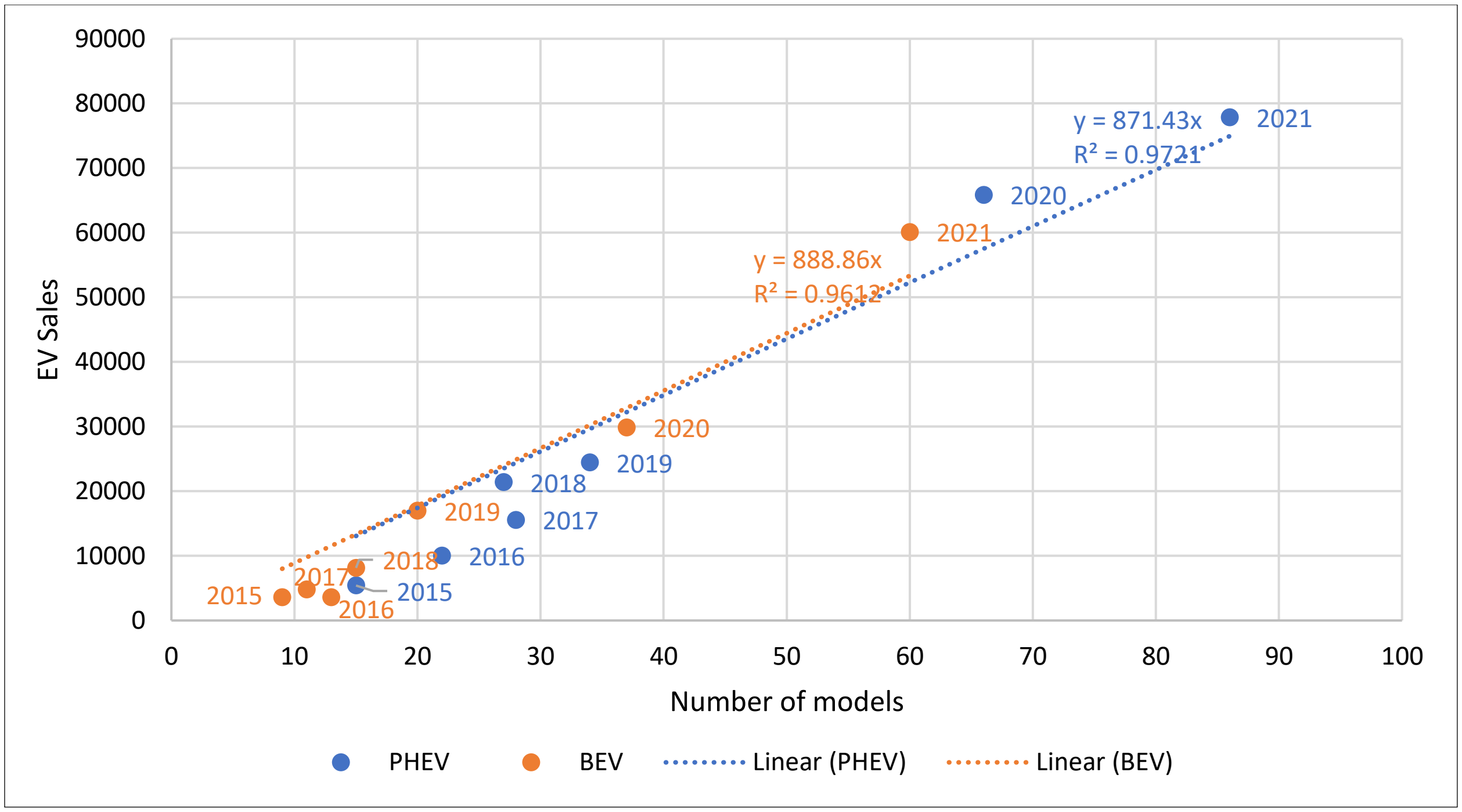

**Figure 17: EV sales vs model availability sensitivity in Sweden, 2015-21**

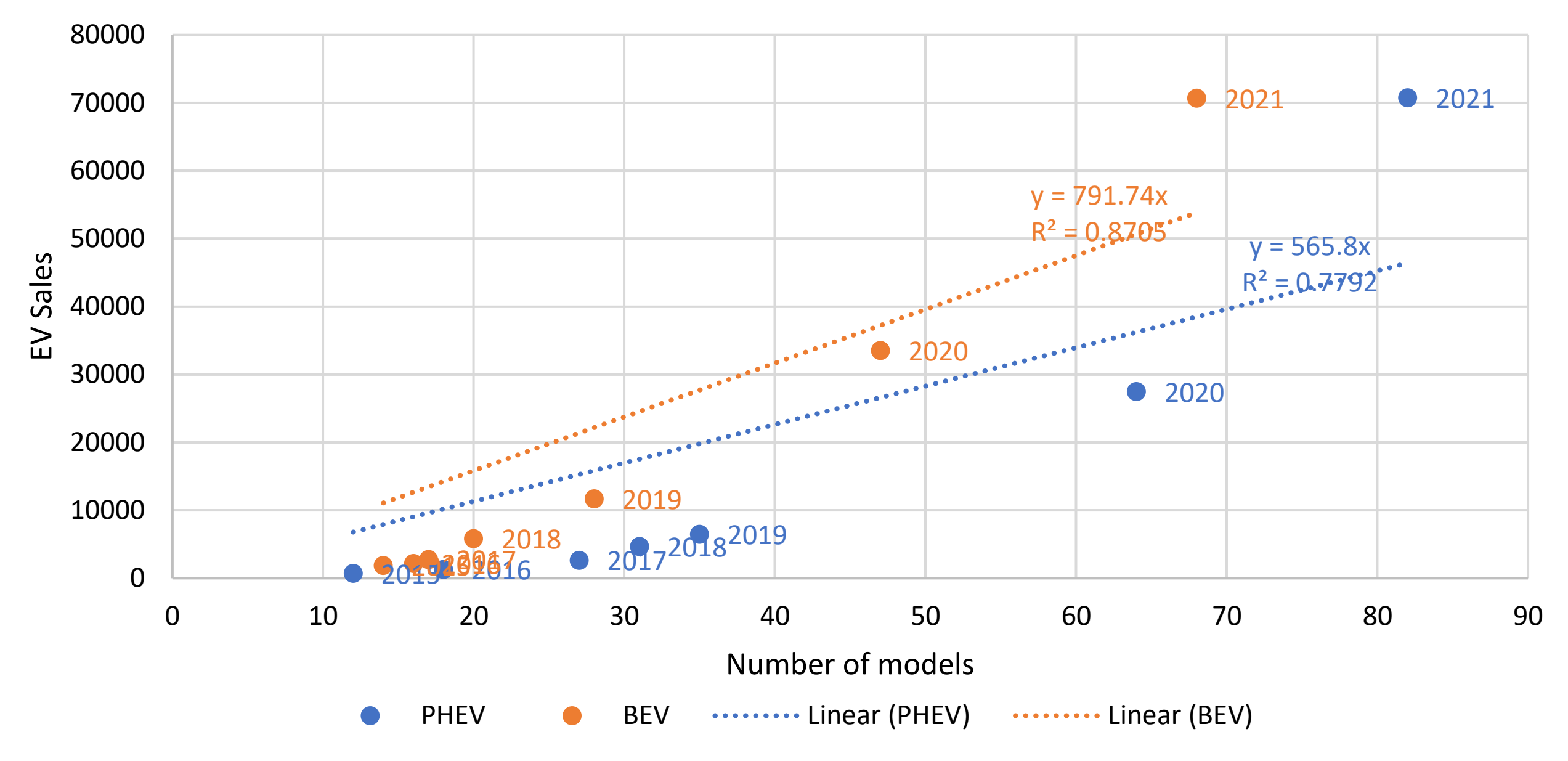

**Figure 18: EV sales vs model availability sensitivity in Italy, 2015-21**

To further understand the distribution of vehicle registrations and consumer choices in the context of the feebate mechanisms, we analyze vehicle registrations and vehicle parameters data from the European Environment Agency (EEA) database for the year 2020, as shown in Table 8. We consider NEDC emissions cycle estimates for this analysis, as not all reporting had transitioned to WLTP estimates in 2020. Based on the average vehicle fleet emissions and the EU emission target in 2020, we look at three emission classes between 0-95 $gCO_2$/km, 96-130 $gCO_2$/km, and greater than 130 $gCO_2$/km.

**Table 8: Country-wise vehicle parameters by emissions class for the year 2020** (European Environment Agency, n.d.)

| ***Year 2020*** | **Avg NEDC** | **% Registrations of total** | **Avg. WLTP** | **Avg. Mass (Kg)** | **Avg. Engine Capacity (cm3)** |
|---|---|---|---|---|---|
| France | 0-95 | 28% | 75.8 | 1398 | 1481 |
| | 96-130 | 66% | 135.3 | 1315 | 1330 |
| | > 130 | 6% | 175.9 | 1678 | 1935 |
| Germany | 0-95 | 19% | 45.6 | 1629 | 1632 |
| | 96-130 | 50% | 137.8 | 1367 | 1452 |
| | > 130 | 31% | 188.9 | 1749 | 2123 |
| Sweden | 0-95 | 40% | 42.8 | 1799 | 1775 |
| | 96-130 | 35% | 139.2 | 1415 | 1475 |
| | > 130 | 26% | 180.7 | 1759 | 2018 |
| UK | 0-95 | 21% | 64.0 | 1566 | 1498 |
| | 96-130 | 53% | 137.3 | 1354 | 1367 |
| | > 130 | 26% | 187.0 | 1776 | 2068 |
| Italy | 0-95 | 23% | 102.3 | 1239 | 1298 |
| | 96-130 | 63% | 135.2 | 1315 | 1351 |
| | > 130 | 14% | 179.7 | 1695 | 1904 |

As in Table 8, Sweden has the highest share of vehicle registrations in the 0-95 $gCO_2$/km NEDC range, which translates to an average of ~43 $gCO_2$/km (WLTP), which is in line with the dominant PHEV sales. It is interesting to note that in this segment, Sweden has the highest average mass per vehicle (~1800 kg) and average engine capacity (1775 $cm^3$), indicating the dominance of gasoline-electric hybrids. As shown in Figure 6, the Swedish $CO_2$ tax is relatively low, compared to the rebates being offered, thus, not serving as a strong disincentive for larger cars.

In the case of France, the 96-130 $gCO_2$ range has about 66% of the vehicle registrations in 2020, which has remained relatively the same when compared to the 2015 data at NEDC test levels. In this segment, while the average vehicle mass has increased by about 2% between 2015-20, the average engine capacity has declined by about 6.5%.

In Figure 19, we can see the change in average $CO_2$ emissions of all new passenger cars sold in each year between 2010 – 19 in France relative to EV sales. It can be observed that although EV sales continued to increase, the average passenger car emissions increase between 2016 – 18. Also, during the same time, the coverage of the malus scheme was extended from about 15% of all vehicles sold in 2016 to about 38% of all vehicles sold in 2019. A logarithmic curve fit for EV sales and average new car emissions provides the best trendline fit based on R-squared values, indicating that the early benefits of increasing EV sales saw a rapid decline in average emissions, and going forward, France will require a significant increase in EV sales to achieve substantial emission reduction benefits. This is likely a consequence of the EU $CO_2$ standards design, which was typically set every 5-years, while feebate revisions are not aligned in the same timeline.

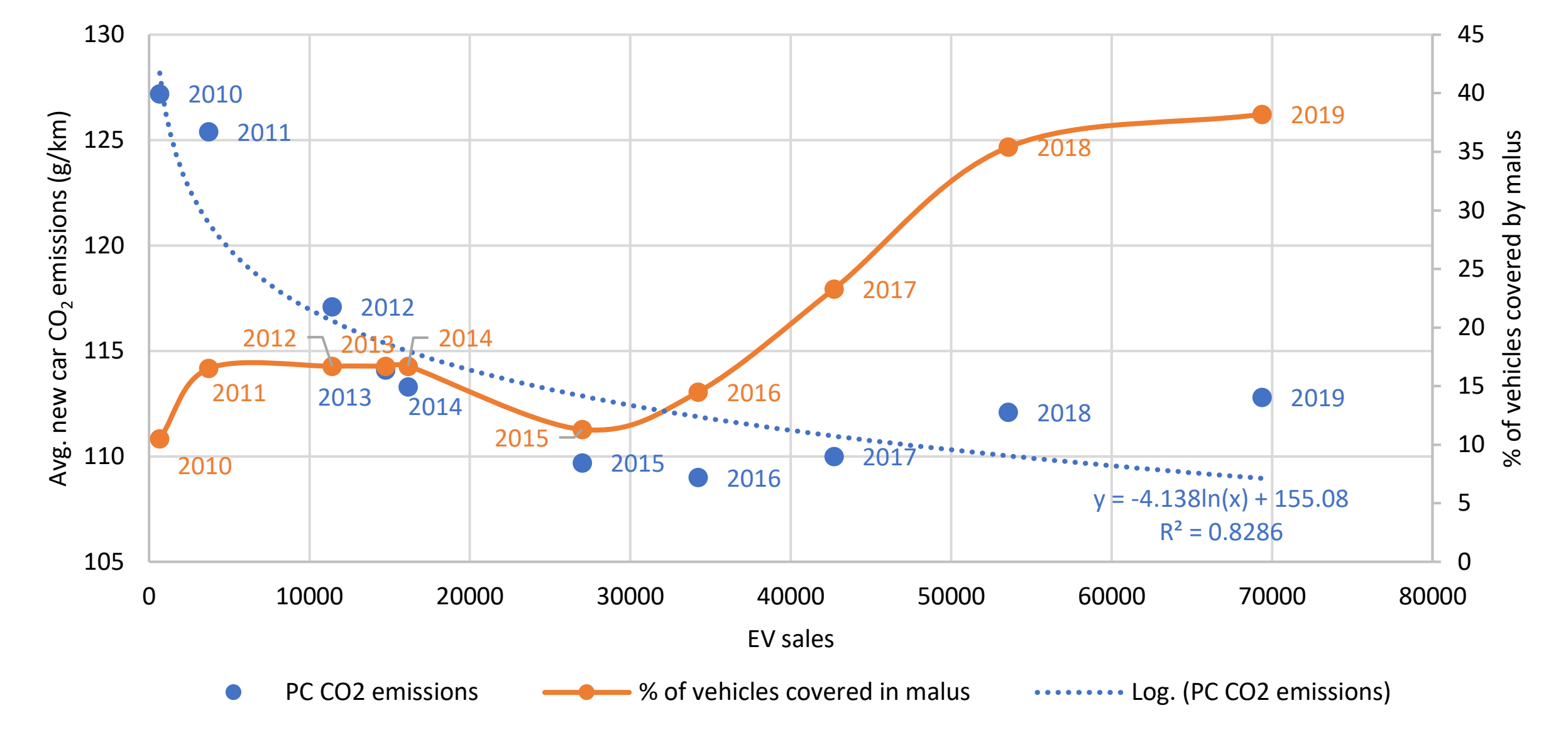

**Figure 19: EV sales vs average passenger car $CO_2$ emissions (g/km, NEDC) in France, 2010-19**

As can be seen in Table 8, the average WLTP-based emission value for over two-thirds of the vehicles registered in France is 135 gCO2/km, which is at the lower end of the malus, as the donut hole extends till 132 gCO2/km. In France's case, the higher bracket of emissions has only 6% of vehicles registered, and thus will require the government to consider policy measures that will enhance emission-based taxation on the median group of vehicle buyers, while also incentivizing automotive OEMs to manufacture vehicles within certain price caps to make ZEVs more affordable and accessible to a larger share of buyers.

Italy has a similar distribution to France, with two-thirds of the registration in the 96-130 $gCO_2/km$ category and having similar average WLTP emission values. **Italy would likely need to consider a similar policy to France but will also have to re-calibrate its malus curve to be more stringent** and include more vehicles within its scope, as the French experience has shown over the last decade.

**The UK, Germany, and Sweden, all have almost one-third of vehicle registrations in the highest emissions bracket (>130 g$CO_2$/km-NEDC) and could likely have a significant impact on EV adoption if they raised the emission-based taxes sufficiently for all vehicles with WLTP emissions above at least 180 g$CO_2$/km**, as a starting measure, and then, gradually increase the fee rate for all vehicles. Germany has among the highest weighted average engine capacity across countries, and it will be interesting to see how the engine displacement-based taxation, in addition to the $CO_2$ tax, will play a role in defining consumer purchase behavior.

Figures 20-23 provide interesting insights on how countries have structured the "malus" component of the feebate mechanisms, based on their 2020 vehicle registration data (European Environment Agency, n.d.). We plot the share of registrations by emission class as defined in each country's emission tax regulation, and the secondary axis plots the emission tax for the mid-point in each emission class (with the highest emission class measured up to 300 g$CO_2$/km). Sweden and Germany are the only two countries that have most registrations within the coverage of the emission fee. In the UK, the slope of the fee curve becomes much steeper after the 150 g$CO_2$/km value, covering over two-thirds of the registrations below that. In Italy, the malus curve only covered 13% of vehicle registrations in 2020.

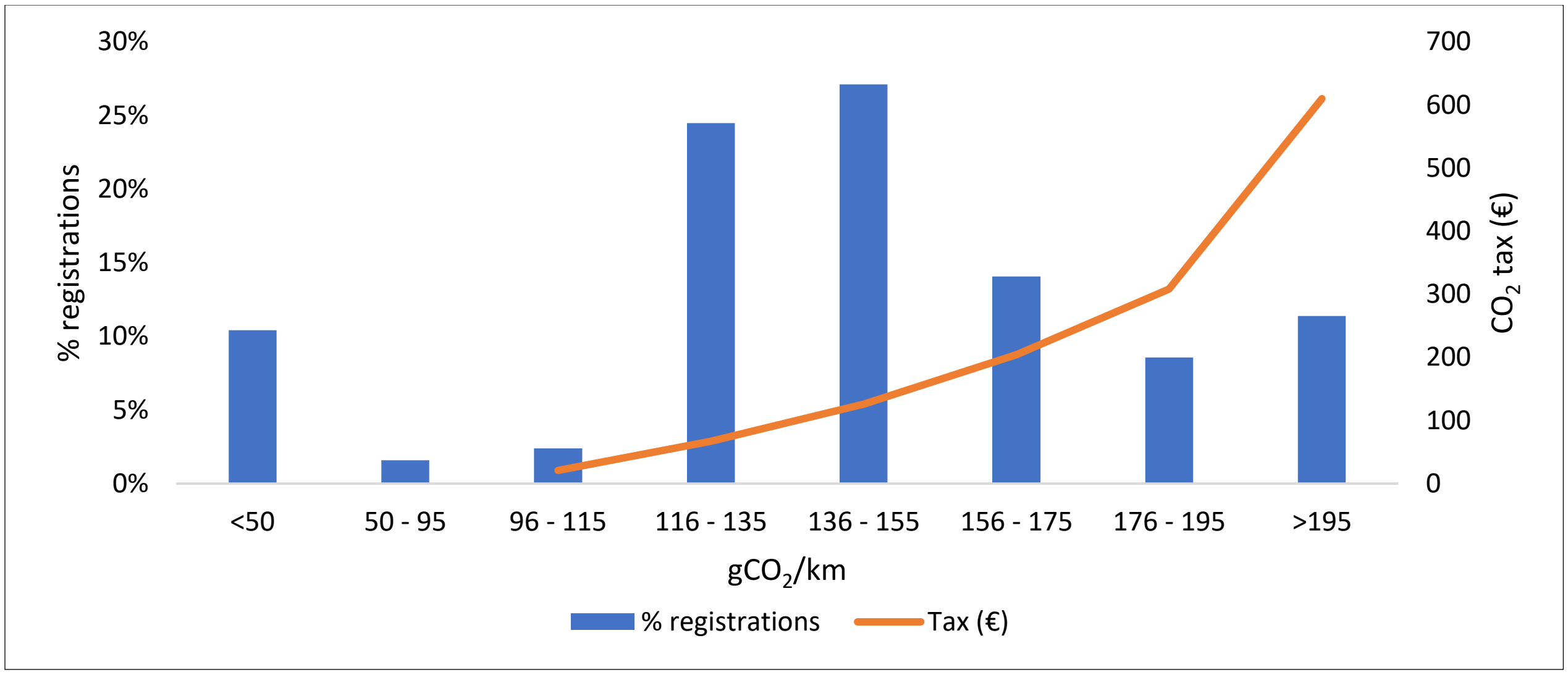

**Figure 20: Share of vehicle registrations by emission categories in Germany (2020)**

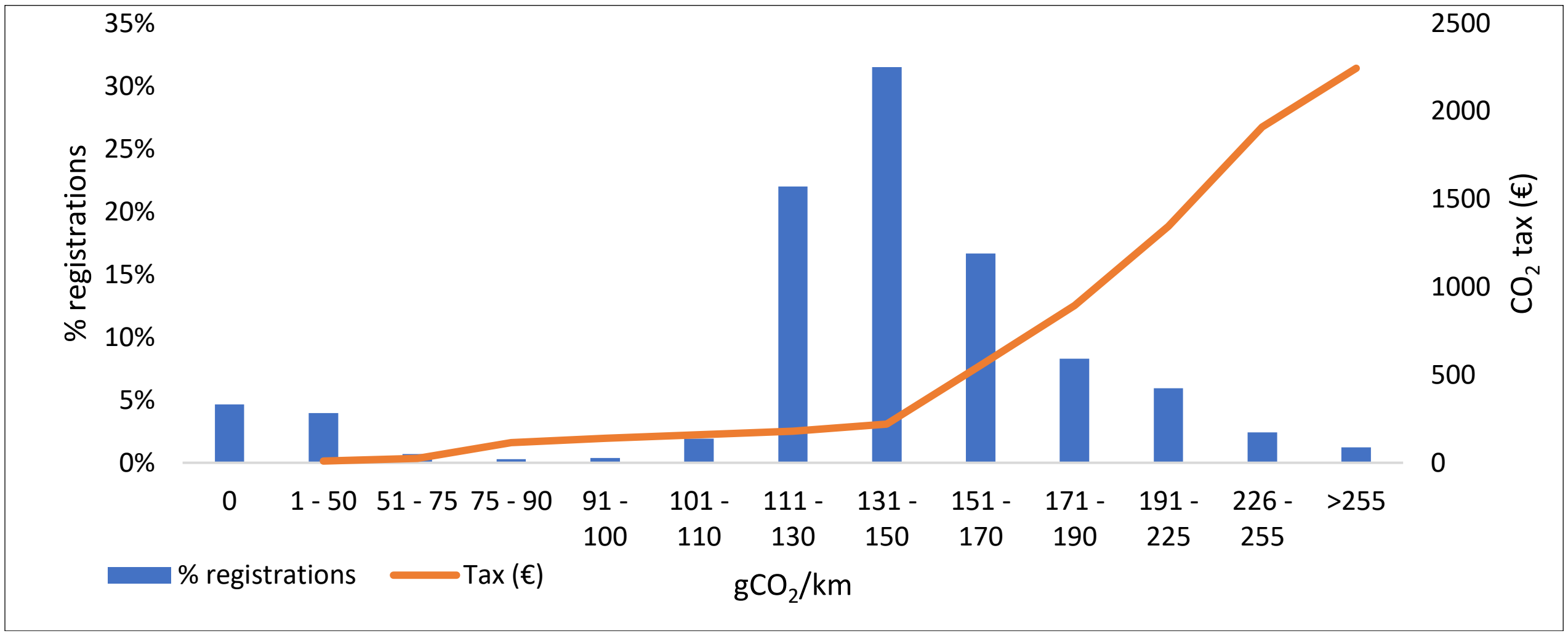

**Figure 21: Share of vehicle registrations by emission categories in the UK (2020)**

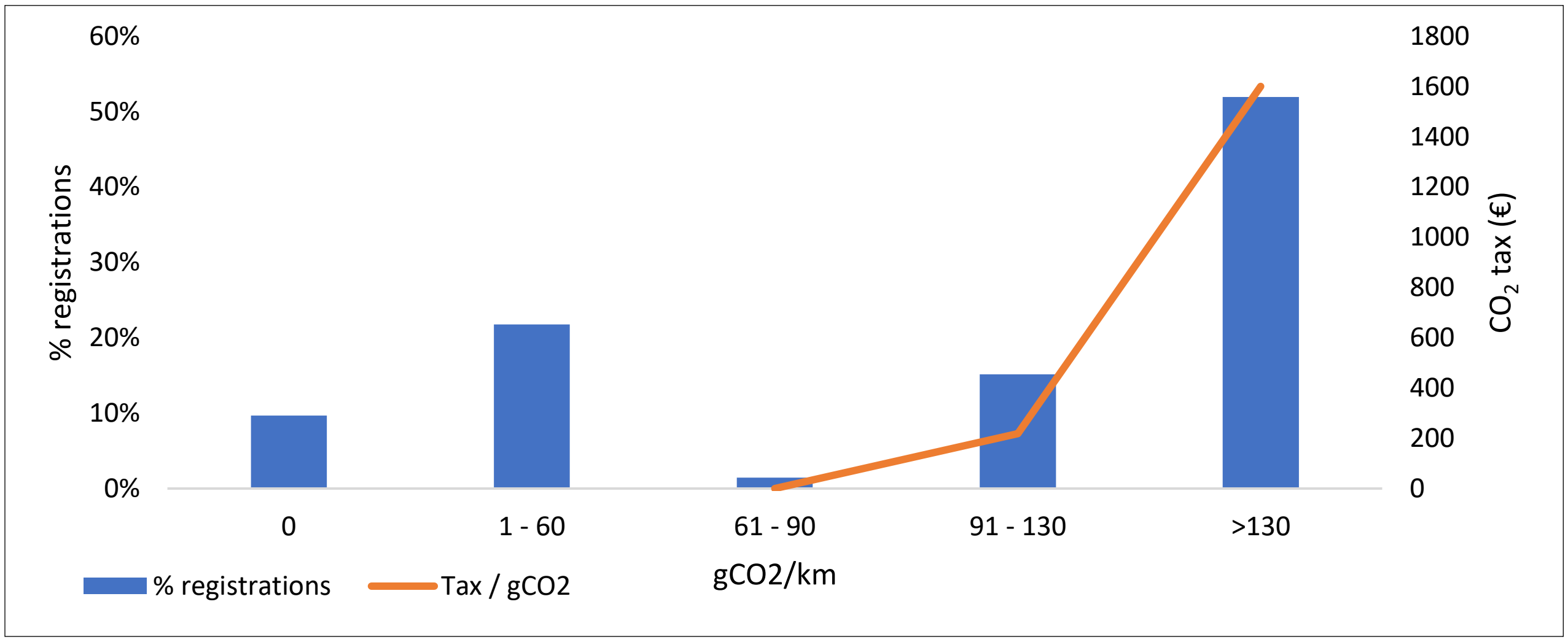

**Figure 22: Share of vehicle registrations by emission categories in Sweden (2020)**

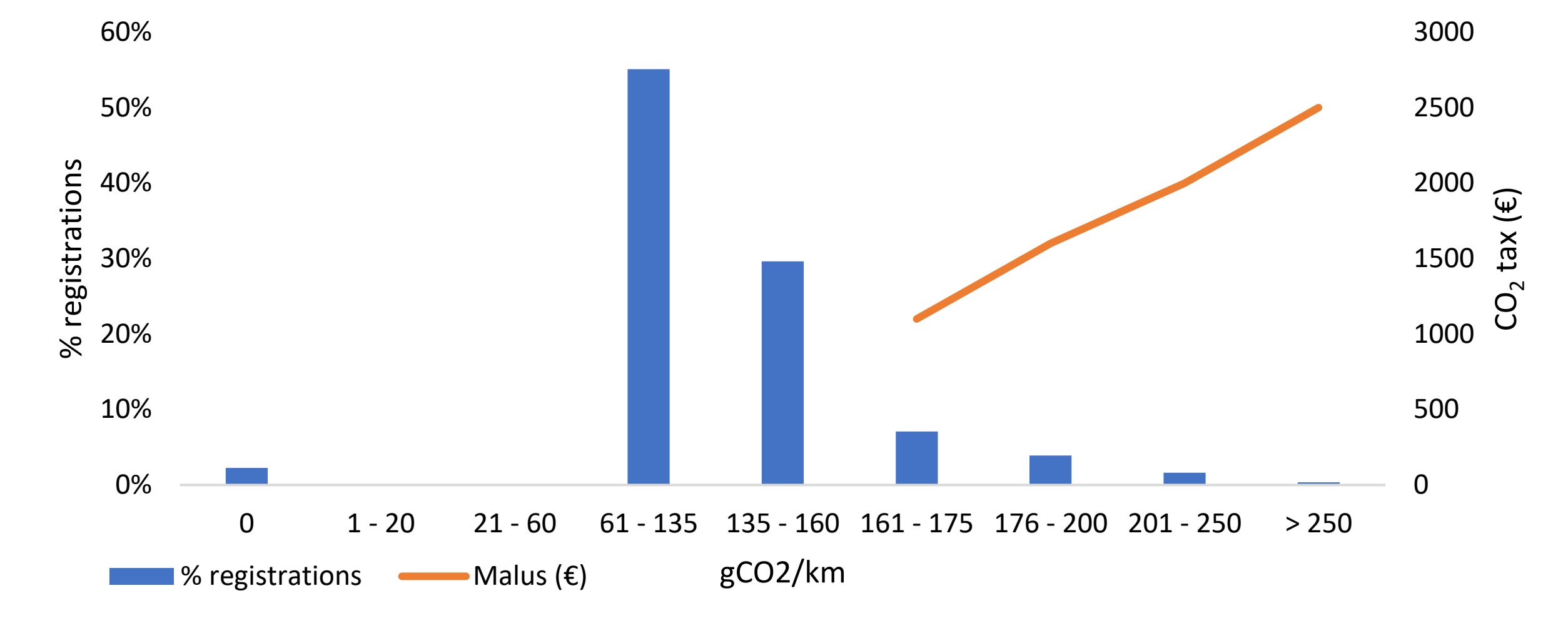

**Figure 23: Share of vehicle registrations by emission categories in Italy (2020)**

## 5. Other Considerations in Designing a Feebate System

While the above analysis across countries provides an insight into average fleet emission values, its impact on vehicle registrations as well as EV model availability and sales, it also sets the base to understand consumer choices in more detail, specifically in the context of prevailing feebate mechanisms.

### 5.1 Preserving consumer preferences

The feebate mechanism can be designed to preserve consumer choice by imposing a fee and offering a rebate within the range of vehicle sizes and types preferred by car buyers in any given market. A size-neutral design could still potentially incentivize OEMs to invest and manufacture EVs including ZEVs, that meet the respective emission thresholds, with offerings within each segment based on consumer demand. If there is a CAFE standard with multiple vehicle types, an equivalent feebate can have separate functions for each vehicle type (Gillingham, 2013), although, a feebate system that treats all vehicles equitably, without any attribute adjustments across different categories would be ideal (German and Meszler, 2010; Kim, n.d.). Challenges of a size-neutral design would most likely include differential impacts on automakers based on vehicle portfolios, although Greene et al., in their study find that feebates can lead to enhanced manufacturer revenues, given the higher value-add for new-technology vehicles (David L. Greene et al., 2005).

The French government in 2007 underestimated the response to the feebate scheme. The demand for smaller and more fuel-efficient cars in the early years and subsequently, for EVs, rose rapidly. While the fees were increased over years, the bonus eligibility included a price-cap on EVs (as in most countries for EV incentive eligibility), thus, in part, forcing automakers towards mid-sized, cost-effective EVs. But given the urgency for a shift to EVs, it was imperative that the slope of the fee line be adjusted to accelerate EV adoption. With a sharp increase in fees from January 2020 to €20,000 (for >184 gCO2/km) and €30,000 from January 2021, from a base of €12,500 in 2019, EV sales saw a significant increase in France from a 3.1% share in 2019 of new vehicle sales to a 15.4% share in 2021. During this time, they also extended the 2020 purchase bonus to 2021 instead of phasing it down, keeping demand robust during the pandemic year, in an otherwise subdued automotive sales environment.

We investigate consumer preferences for vehicle segments across passenger cars and SUVs as can be seen in Figures 24-26 (for this analysis, MPVs are included within the SUV segment). Essentially, we see a general shift to D-segment EVs among cars in most countries, except Italy, where there is a clear shift to A-segment EVs. We can observe similar emerging trends in the case of SUVs, with all countries moving to C-segment EVs, with France, Germany and Italy moving upwards in size, while Sweden and UK move downwards in size.

Across all vehicle type classifications for EVs (cars and SUVs), we can see that France and Germany have transitioned in a similar manner, while Sweden and the UK have shown a similar transition.

If we compare the EV consumer choice trends to overall LDV sales in these countries, we find similar trends, wherein LDV sales have essentially been dominated by C-segment in most countries. Germany, Sweden, and the UK are all dominant C-segment markets for LDV sales, while Italy and France are dominant B-segment markets (Figure 27).

**Two key insights evolve:** (i) between 2015-2021, all five countries have seen an overall convergence to C-segment EVs being the dominant share; and, (ii) in the early years of the transition, consumer choices were probably constrained by affordability and model choices, but gradually over time, with greater model availability as well as significant changes to the $CO_2$-based taxation, consumers are tending to preserve their choices, and not necessarily change vehicle size preferences.

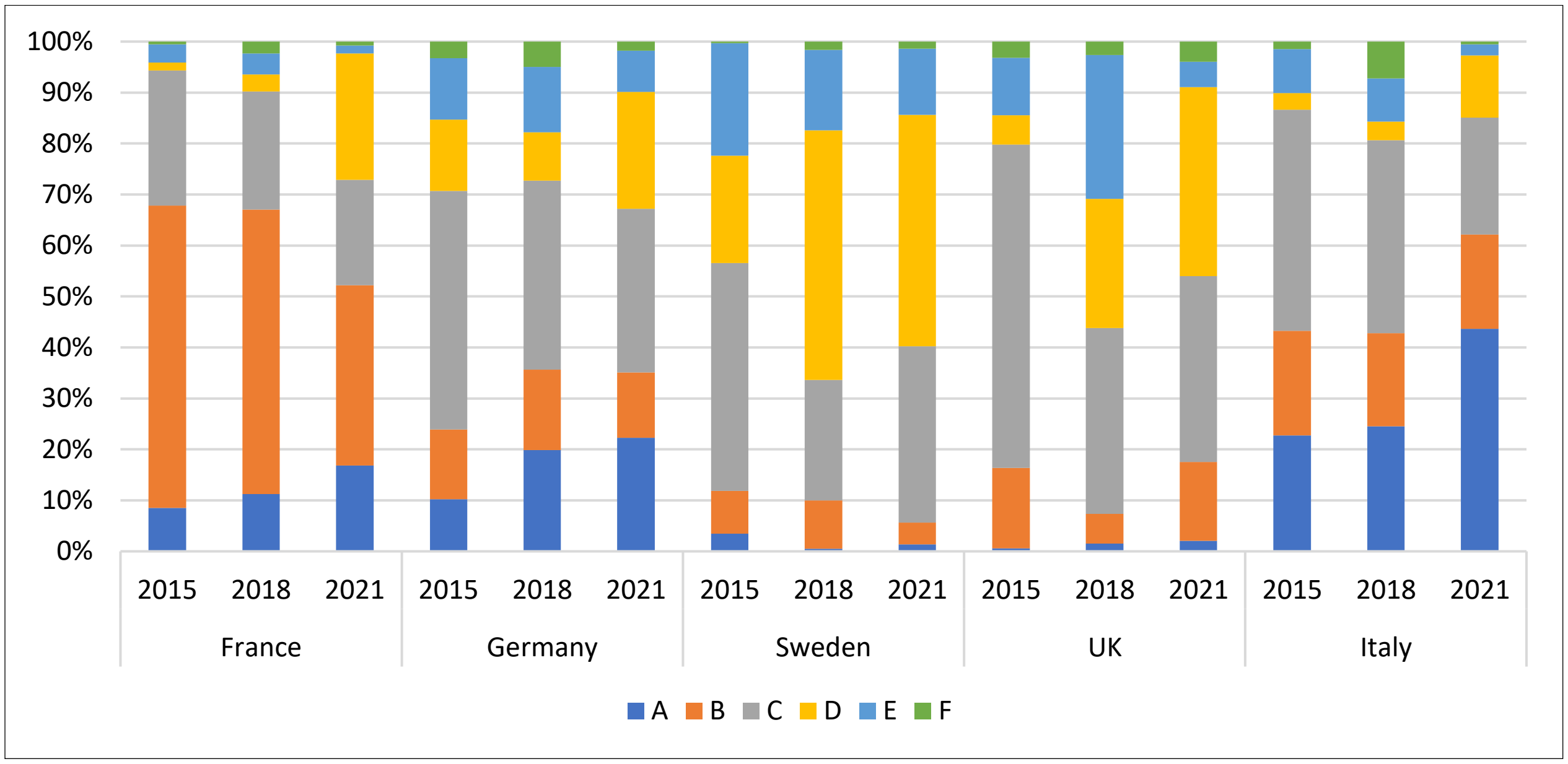

**Figure 24: Share of EV car sales by segment size**

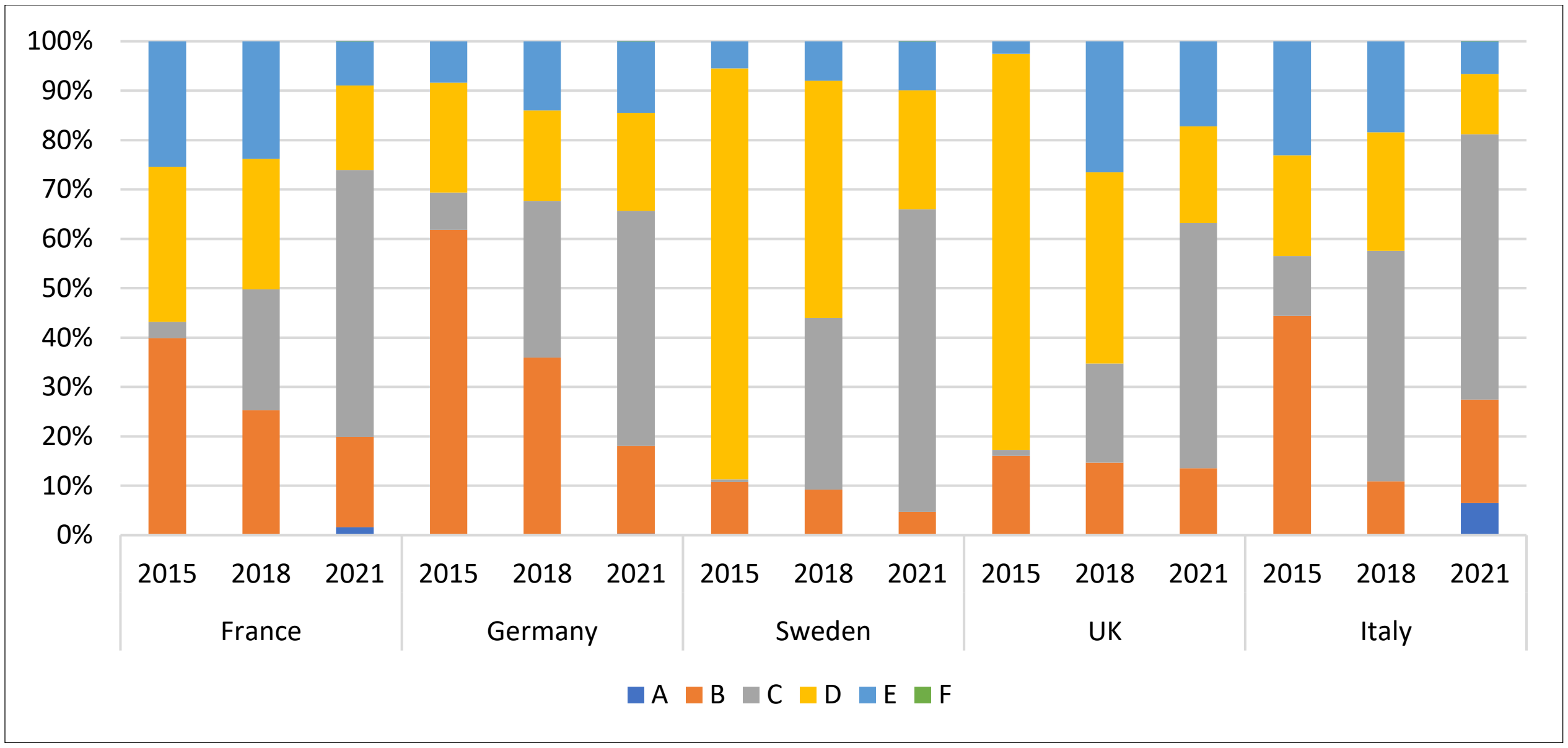

**Figure 25: Share of EV MPV/SUV sales by segment size**

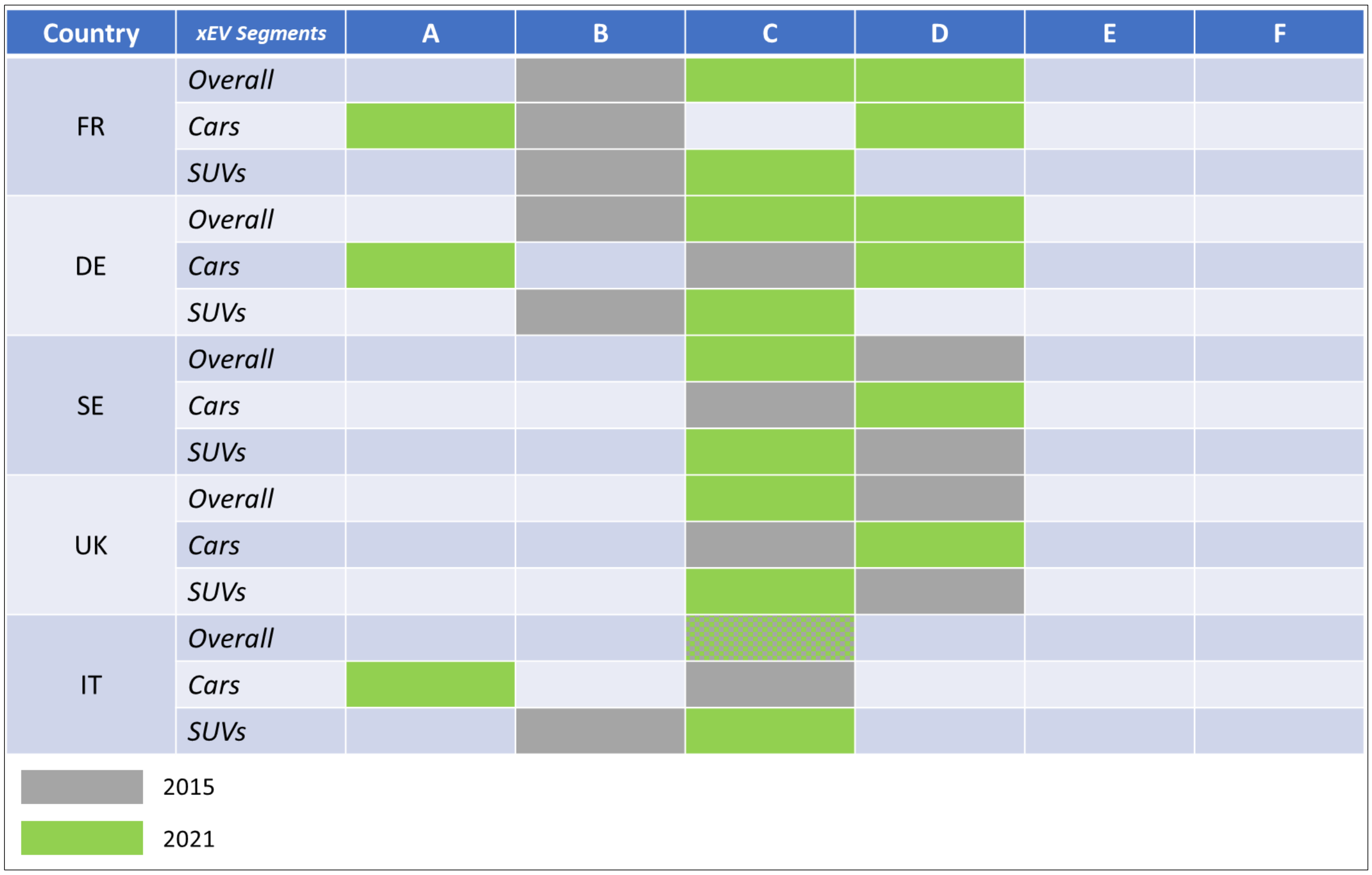

**Figure 26: Transition of predominant vehicle segment choices for EVs across vehicle types, 2015-21**

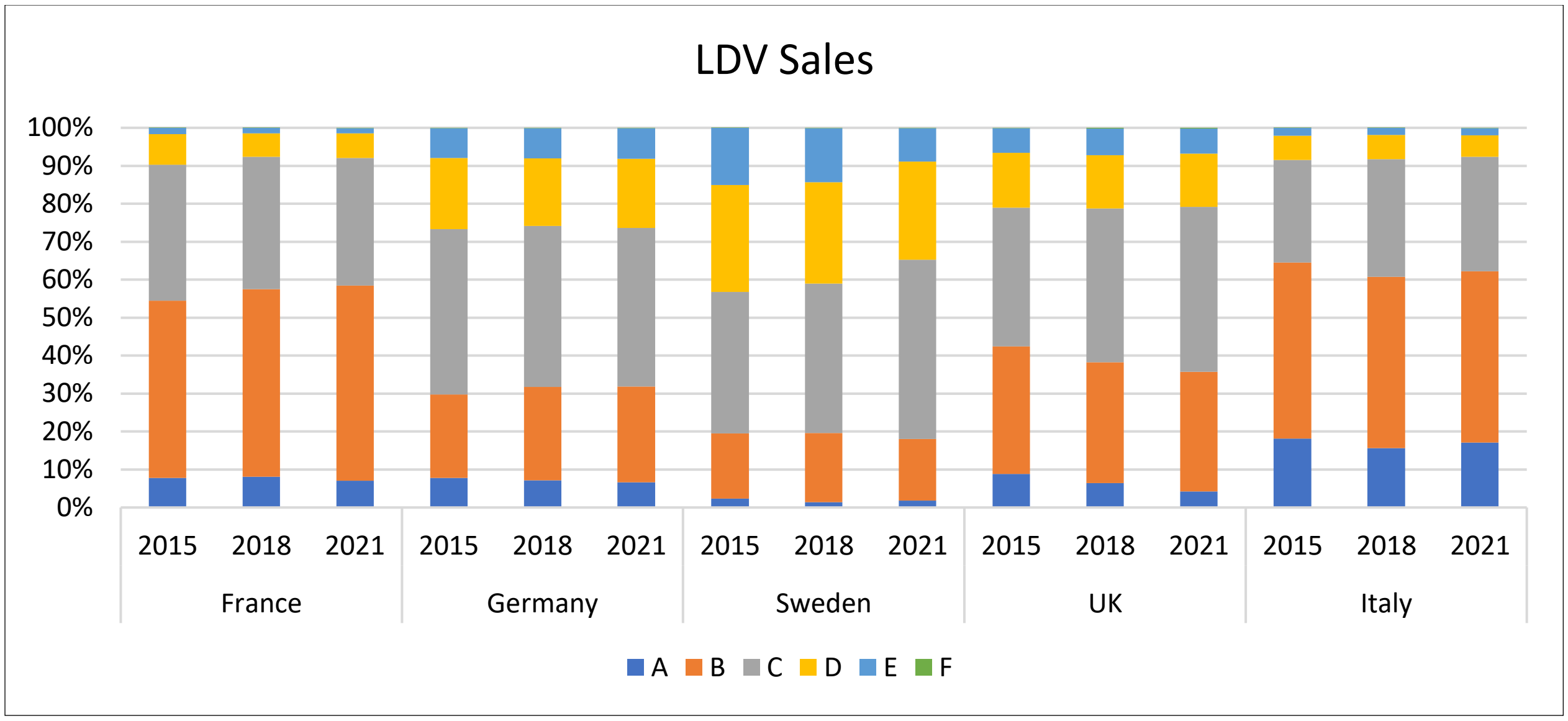

**Figure 27: Transition of predominant vehicle segment choices for total LDV sales, 2015-21**

## 5.2 Political acceptability and revenue-neutrality

Legislative representatives often make an argument that a feebate is viewed as a tax and may not be politically feasible. At the same time, unlike general tax revenue which is often appropriated by congressional or legislative approved budgets annually, a specific feebate for clean vehicles would almost instantly redistribute the tax revenue gained from higher emitting vehicles to reward consumers purchasing efficient or low emission vehicles. Given its potentially redistributive nature, a feebate need not necessarily be seen as a tax, but rather a carbon dividend payout to society (Ramseur and Leggett, 2019). Feebates can also have strong public support if they are deemed to be fair, which is where the lessons for an effective feebate design play a critical role (Martin et al., 2014).

Another important feature of feebates is that of revenue-neutrality. They can be designed such that the fees levied can at least be equal to the rebates offered plus the administrative costs, through setting the pivot point with annual revisions based on past consumer responses and market trends, ensuring that the program stays revenue neutral.

From a fiscal perspective, the revenue generating aspect of the feebate structure has worked for countries including France and Sweden. It took France a few years to forecast and manage the feebate to balance the flow of revenues, achieving surplus revenue since 2014. As seen from the French experience, feebates can be cost-effective and award both firm-level and household-level responses to reducing emissions. In Sweden, estimates by the Swedish Transport Agency expected a surplus of SEK 0.43 billion (~€42 million) in 2018, SEK 0.09 billion (~€9 million) in 2019 and SEK 0.58 billion (~€56 million) in 2020 (Ministry of Finance, 2017). In Germany, given the hybrid nature of the policy, it is yet to be analyzed how much of the revenue collection will help co-finance the EV rebates. As per the German government, €2.09 billion have been earmarked from 2020 onwards to fund the EV rebates program, at least till 2025 ("So funktioniert der neue Umweltbonus," n.d.). In Italy, the government had allocated €60 million for 2019 and €70 million for 2020 and 2021, for rebates towards BEVs and PHEVs. But given that the Italian bonus-malus scheme was launched in 2019, the revenue flows are yet to be analyzed in detail.

### 5.3 Ensuring equity in the EV transition

One of the key elements of achieving the EV transition, and even more so in case of a ZEV transition, will be its focus on equity. It will be critical to make EVs more affordable to all middle- and lower-income consumers, to achieve a scaled and effective transition in the next decade. Various geographies have made efforts to address some aspects of equity, by way of additional rebates for EV purchases by low-income households, used-EV schemes, or vehicle trade-in programs (California Air Resources Board, n.d.; Wallbox, n.d.; Wappelhorst, 2021).

For 2021, France provided a maximum combined rebate of € 12,000 for the purchase of used or new BEVs and PHEVs for scrapping an older ICE vehicle (older than 2011 for diesel and older than 2006 for gasoline), subject to the reference tax income being lower than € 13,489 (for the year 2021). They also have a low emission zone bonus of € 1,000 if an individual lives or works in a low emission zone, which can be of benefit to some low-income families who are either residing in such areas or beneficiaries of affordable housing in low-emission zones or employed in such areas. The low-income rebate conditionality also distinguishes between vehicle buyers as 'average commuters' or 'heavy drivers', based on their home-to-work commute distance.

Germany does not have a specific low-income grant for purchase of EVs but provides incentives on purchase of used EVs. Although, the conditionality for receiving a rebate on purchase of a second-hand EV is that the vehicle should not have received the federal rebate when it was originally purchased (Wallbox, n.d.). This condition potentially acts as a barrier for low-income households to access EVs. Alternatively, the government could consider a program wherein the rebate on a used-EV can be claimed when it is sold, such that the vehicle is at least 1 or 2 years old.

In Italy, the bonus-malus scheme includes a provision for low-income households, i.e., those households with an Equivalent Economic Situation Indicator (ISEE) of less than € 30,000 (for 2021). Such households who purchase new EVs with a power of less than 150kW and a list price less than € 30,000 (excluding VAT), will be provided 40% of expenses incurred, and this scheme cannot be combined with any other eligible benefits (European Automobile Manufacturers' Association, 2021).

The UK illustrates one example of potential equity impacts arising out of changes to the feebate design. In keeping with conventional wisdom which holds that increasing vehicle emission-based taxes should lead to higher EV sales, the UK revised its vehicle taxation norms in 2020, and witnessed an EV sales increase of about 2.3 times over 2019, driven largely by BEVs. But at the same time, in 2021, the UK reduced the EV purchase rebate to GBP 1,500 or approximately € 1,800. A disproportionate reduction in rebates could have a negative impact on both the industry and consumers, bringing forth equity and affordability concerns for many.

To be able to effectively manage the transition, it will be imperative that policies are designed in a manner that helps shift middle- and higher-income households to ZEVs while lower income households can move from ICE-powered vehicles to PHEVs or to smaller BEVs as potential starting points. As the scale benefits for ZEVs are realized in the market, they will lead to an overall reduction in vehicle prices and a better realization of resale values for ZEVs, improving accessibility and affordability for consumers on the lower end of the income spectrum (Koehler et al., 2019).

## 6. Key insights and considerations of a feebate design for a ZEV transition

For all countries, an EV transition poses multiple challenges, foremost among them being financial resources, industry shifts and consumer behavior. As seen in earlier sections, various EU countries continue to rely on feebates as an effective policy tool to achieve transitions towards low and zero emission vehicles.

Sweden and Germany, both markets with significant PHEV sales in the past few years, serve as a good example to contextualize the role of feebates to move from an EV to a ZEV transition. With a relatively small 'donut-hole' and offering graded rebates (a continuous rebate function) for all vehicles emitting below 60 $gCO_2$/km, the Swedish feebate provides incentives for both BEVs and PHEVs and remains a dominant PHEV market as of 2021. On the other hand, Germany witnessed a 3X and 4.5X increase in BEV and PHEV sales, respectively, in 2020 compared to 2019; and, about a 1.7X increase in both BEV and PHEV sales in 2021, relative to 2020. It is yet to be established whether the revised emission-based tax will have an accelerated impact on ZEV sales in the near future or if Germany would have to revise the feebate to incentivize a shift to BEVs and other ZEV alternatives. The German policy, while transitioning to a partial feebate model, has yet to find a balance in terms of the fee and the rebates offered. The tax implications with the revised feebate are still quite low, being a maximum of additional €140 for vehicles with emissions up to 195 $gCO_2$/km (as of 2021).

While there are key elements of a feebate mechanism that should be considered while designing it for implementation, there is no one single or 'optimal' design. There are different policy objectives that could be served by a feebate mechanism, which would influence its design and its effectiveness as a policy solution. We highlight few policy objectives where in feebates can play a role in meeting those goals, and possible conditions under which they would work.

One of the most fundamental policy objectives could be to shift to more fuel-efficient ICE vehicles. Italy is a good example, which essentially imposed a fee on less than 15% of the vehicles sold (among the highest emission classes). The possible reasons for a sub-optimal design include protecting Italy's domestic automotive industry, as well as its own domestic political and policy narrative regarding climate change. At a global level, using feebates to shift to more fuel-efficient vehicles could be well-served in countries (especially developing economies) where technology leapfrogging to EVs is challenging. A country could be a net importer of vehicles or lacks the technical knowledge.

Using feebates to facilitate a shift to EVs has emerged as the more recent objective, especially as countries like Sweden and the UK have looked to leverage this policy choice in recent years. At the same time, it should be noted that feebates can be amended over time to meet differing policy goals, as in the case of France and Germany. France's initial objective over a decade ago was to use the feebates to shift the market to more fuel-efficient ICE vehicles, and over time has revised the feebate design to facilitate a shift to EVs, and in recent years, a shift to ZEVs. A more constrained version of the policy objective would be to use feebates to shift to ZEVs. Countries like India, which are hesitant set national EV targets but at the same time only incentivize a shift to BEVs (and no PHEVs), could benefit from a feebate mechanism designed to meet the ZEV transition objective. Moreover, the financial sustainability of stand-alone EV incentive programs has come into question, and thus, a self-financing market mechanism could be the 'need of the hour' solution (France was able to achieve surplus revenue from its feebate within 4 years of implementing it).

Irrespective of the policy goals, a feebate will impact both the supply side, i.e., the automotive industry and the consumer side. In either case, the cost of an ICE vehicle in most segments becomes prohibitive for sale (industry viewpoint) or purchase (consumer viewpoint). Depending on the feebate design, the industry will shift its strategy towards more policy-compliant vehicles, likely to increase model availability and bring down technology costs. Given that the fee is effectively imposed on consumers, the mechanism has almost a parallel effect on vehicle purchase decisions, almost creating an equilibrium of sorts in the market between supply and demand objectives. Feebates could also be used to address additional policy goals, such as curtailing growth of certain kinds of vehicles, for example, very large SUVs.

**Why do we need a ZEV transition?** The single largest objective would be to maximize the benefits of an EV transition in terms of GHG emissions reduction. Feebates can help us achieve that objective but will require design considerations to promote a shift to ZEVs.

While the current focus of feebate mechanisms has been to incentivize both BEV and PHEV sales, certain aspects of the mechanism will need a different approach to push for a ZEV transition, while addressing any potential equity concerns, for both producers and especially, consumers. Based on a review of the feebate mechanisms prevalent in major European countries as presented in this paper, **we highlight thirteen key insights towards designing a feebate policy that can facilitate an inclusive ZEV transition in the next decade** (Figure 28).

**First and foremost, while feebates provide a certain source of revenue, the policy context within which the feebate is made effective will be critical.** Feebates can be classified as pure or partial, based on whether the "fee" and "rebate" parts of the mechanism were introduced under one mechanism (example, France and Sweden) or instituted separately (example, Germany and the UK). Further, in cases where the fee revenue generated is not mandated to finance rebates either by regulation or legislation, it could lead to sub-optimal resource allocations towards EV purchase incentives and charging infrastructure (for example, the UK). A clear mandate of funds utilization provides a certainty to the market and provides the government flexibility in planning for additional budgetary allocations to bridge the resource gap.

**Second, identifying the distribution of vehicles sold by emissions ($gCO_2$/km) is a key first step in the feebate design strategy**, followed by an understanding of the vehicle prices within each emission class, and then choosing the feebate design, i.e., the functional form, the efficiency parameter, and the pivot point. Further, equity can also be addressed within the feebate mechanism by developing a better understanding of household income levels and prices of vehicles purchased, especially, the vehicle purchases made by the $50^{th}$ percentile of households in income terms, prior to deciding the functional form, slope of the curve and pivot point.

**Third, focusing on a single fee parameter, i.e., CO2 emissions, can be a simple yet effective mechanism.** Countries are using combinations of $CO_2$ emissions and other vehicle attributes such as vehicle weight (ex. France), length, and engine capacity (ex. Germany) to estimate the total fees payable on purchase of a new vehicle. Where introducing emission taxes can be a challenge, there may be a case for attribute-based taxation as an alternate measure, but it may still not lead to a technology shift in terms of a transition to EVs, and even more so for ZEVs, as attribute-based taxation may not serve as a strong hedonic pricing mechanism for emissions externalities.

A $CO_2$ emissions-based fee mechanism should form the basis of vehicle taxes, as it will provide manufacturers the flexibility around adjusting other vehicle attributes if they meet the emissions reduction targets. Adjustments can be made in the emissions fee itself rather than introducing additional attributes and fee mechanisms. Although Italy-type markets wherein the emission tax itself is designed to have limited coverage of vehicles in the market, attribute-based fees in addition to emission fees can play a key role in the total effect of disincentives, as well as to gather more political acceptability.

Vehicle attributes are increasingly becoming complex variables with non-linear relationships to vehicle emissions. We have seen that all countries analyzed in the paper have shown an increase in average vehicle mass but have also seen a reduction in average emissions. Once EV sales reach a certain volume and basic taxes are reintroduced on EVs, vehicle weight taxation could have perverse implications due to battery size and weight, attributes of EVs which are still evolving with limited evidence on future trajectory. Similarly, average engine displacement has reduced across all countries, although vehicle footprint has increased. Germany which uses engine displacement fees as well in the feebate design, has seen the lowest reduction in power to weight ratio among the five countries. Attribute fees, such as vehicle footprint-based taxation can have additional benefits in terms of urban design, parking constraints, and road capacity.

**Fourth, the functional form for the fee and rebate should be carefully considered.** A continuous function (preferably a non-linear fee function, with a steep rise in fees for higher emission values) for levying fees (and providing rebates) for every unit increase in $CO_2$ emissions (g/km), rather than a step function, provides the best way to avoid system gaming and ensure continuous incentives to build and purchase lower $CO_2$ vehicles. A piece-wise linear fee function can also be designed to generate sufficient revenues by imposing the highest tax burden on high-emission vehicle buyers as compared to the middle $50^{th}$ percentile of vehicle buyers.

Further, in case of rebates, it is seen that a stepwise function (as compared to the fee function) will likely be more efficient as it can be structured to incentivize PHEVs with higher all-electric range requirements in the interim and target greater rebates towards ZEVs.

**Fifth, periodic revisions in the slope of the curve and the pivot point can help ensure a revenue-neutral system.** Further, providing a two-year horizon on the functional form for the fee gives positive market signals, as seen in the French experience. While France provides a two-year horizon of revisions, Denmark has gone a step further with a clear indication of the revisions in the pivot point for the next ten years, with the limit value being reduced by 3.3% annually from 2022-25, and then by 1.1% from 2026-30 (“Applicable rates for registration tax Ministry of Taxation,” n.d.). A relatively shorter frequency of revisions (every 2-3 years) to the feebate design are likely

to result in more favorable policy outcomes, as compared to a long gestation period, as it might result in either under- or over-estimating the potential for technology and market developments. The slope of the fee curve, which will impact both producer and consumer decisions, can be based on various decision parameters, such as upfront purchase price or Total Cost of Ownership (TCO). The basis for deciding the slope of the fee curve will also have implications on the ZEV transition pathway, in terms of the duration to achieve the targets as well as the financial resources that may be required for the same.

**Sixth, the choice of having a single pivot point or a donut-hole should be based on an analysis of the type of vehicles being sold in the market**, possibly a percentile approach based on vehicle prices and/or emissions. This will define which of the vehicles being sold in the market will be taxed or receive a rebate or be excluded from the feebate mechanism altogether (donut-hole). The choice of the pivot point goes together with the prevailing fuel efficiency and emission norms.

**Seventh, the feebate design needs to be supported by external policy choices** such as a vehicle price caps for incentive eligibility for EV purchases and All-Electric Range (AER) requirements for PHEVs. Overall, as a basic principle, the rebates need to reduce over time, while the fees increase, forcing both, automotive manufacturers, and consumers, to reconsider their choices.

**Eighth, we find that the differences in the point of collection of the emissions-based fee for the consumer will likely play a key role in the transition to ZEVs.** A higher one-time fee collected at point of purchase is expected to be more effective than an annual fee, given consumer discounting of future cash flows, and could well be the difference between choosing a PHEV or a

BEV. Similarly, applying rebates at point of sale are likely to be more effective than tax rebates or staggered incentive payments.

While the rebate itself is structured as a step function and accounts for restrictions on maximum vehicle purchase price and minimum certified range requirements (for all EVs), the highest impact on EV adoption is when the rebate is offered on-the-hood or at the time of vehicle purchase. A rebate applied at the point of purchase would incentivize consumers as the benefit is tangible and they do not have to wait for it. Direct monetary benefits at the point of purchase have a stronger influence on consumer choice than annual tax refunds. The fee and rebate can be applied at different points in the transaction value chain. The Swedish feebate pays the bonus six months after the vehicle registration to counter any temptation of a buyer to collect the bonus and then immediately resell the car (International Council on Clean Transportation, n.d.). As the fee is revised, and EV prices decline, the rebate can be adjusted to lower levels, before being potentially phased out, and only the fee aspect remains to completely disincentivize any ICE vehicle sales over a period. Another key incentive offered in many countries has been registration tax exemptions on purchase of EVs, which will eventually lower the tax revenue from vehicle sales once EV sales reach a sizeable share. In case of feebates, vehicle registration taxes can be set for both ICE and EVs to meet tax revenue requirements, to provide viability gap funding to incentivize consumers to purchase EVs.

**Ninth, we find that BEV sales indicate a higher growth sensitivity than PHEVs for every additional model made available in the market, and it could provide an important basis for future feebate design focused on a ZEV transition.**

The question that comes up is once automotive companies start increasing the model availability across EVs, will that be sufficient in driving higher rates of adoption? This will depend on other factors including vehicle prices, (dis)incentives and the feebate mechanism (in this case). We observe for 2021, that Sweden has a lower availability of EV models compared to others but has the highest EV market share among LDV sales for that year. In contrast, the UK, which has the lowest number of EV models sold in 2021, has an EV LDV market share of 14.4%, much higher than the 8.6% EV LDV share in Italy, even though the latter has a higher number of EV models available to consumers.

**Thus, this brings up an important consideration (Tenth) that model availability alone may not be sufficient to drive higher adoption rates of EVs.** It will be important for policy maker to understand the prices and other vehicle attributes of the EVs available in their respective country markets, and where they stand in respect to comparable existing ICE vehicles in the market and prevailing consumer choices.

**Eleventh, in case of the five European countries, we find the general argument that feebates constrain consumer preferences for larger vehicles or are pro-small cars, does not necessarily hold true.** Between 2015-2021, all five countries have seen an overall convergence to C-segment EVs, similar to ICE vehicle choices, without any specific disincentive on vehicle footprint, and rather has been driven by the inherent fee and the market response by both producers and consumers.

**Twelfth, equity considerations for those in low-income groups and other disadvantaged communities will be critical in ensuring a mass transition to EVs, as well as distribution of benefits across society.** Of the five European countries, except France and Italy, we do not find

any explicit support for low-income households to purchase EVs. Various measures can be considered including used-EV schemes and vehicle trade-in programs. Also, low-income, and disadvantaged communities can be supported to transition to PHEVs before a move to BEVs, given constraints of access to mobility.

**Thirteenth, it is important to have a robust monitoring framework.** It is essential to maintain a strong database of vehicle sales, their pricing in each year, emissions portfolio, and so on, to facilitate a periodic revision of the feebate mechanism, to make it self-sustaining. Making realistic forecasts of responses to feebates has been challenging given the lack of literature on relevant elasticities (Berthold, 2019b). This will also have important implications for equity considerations, especially in PHEV dominant markets where sustained and potentially higher incentives will be needed for a ZEV transition. Such markets will need changes in the fee structure to facilitate a shift among majority of consumers, while suitable exemptions are made for lower income households, such as greater incentives for PHEVs.

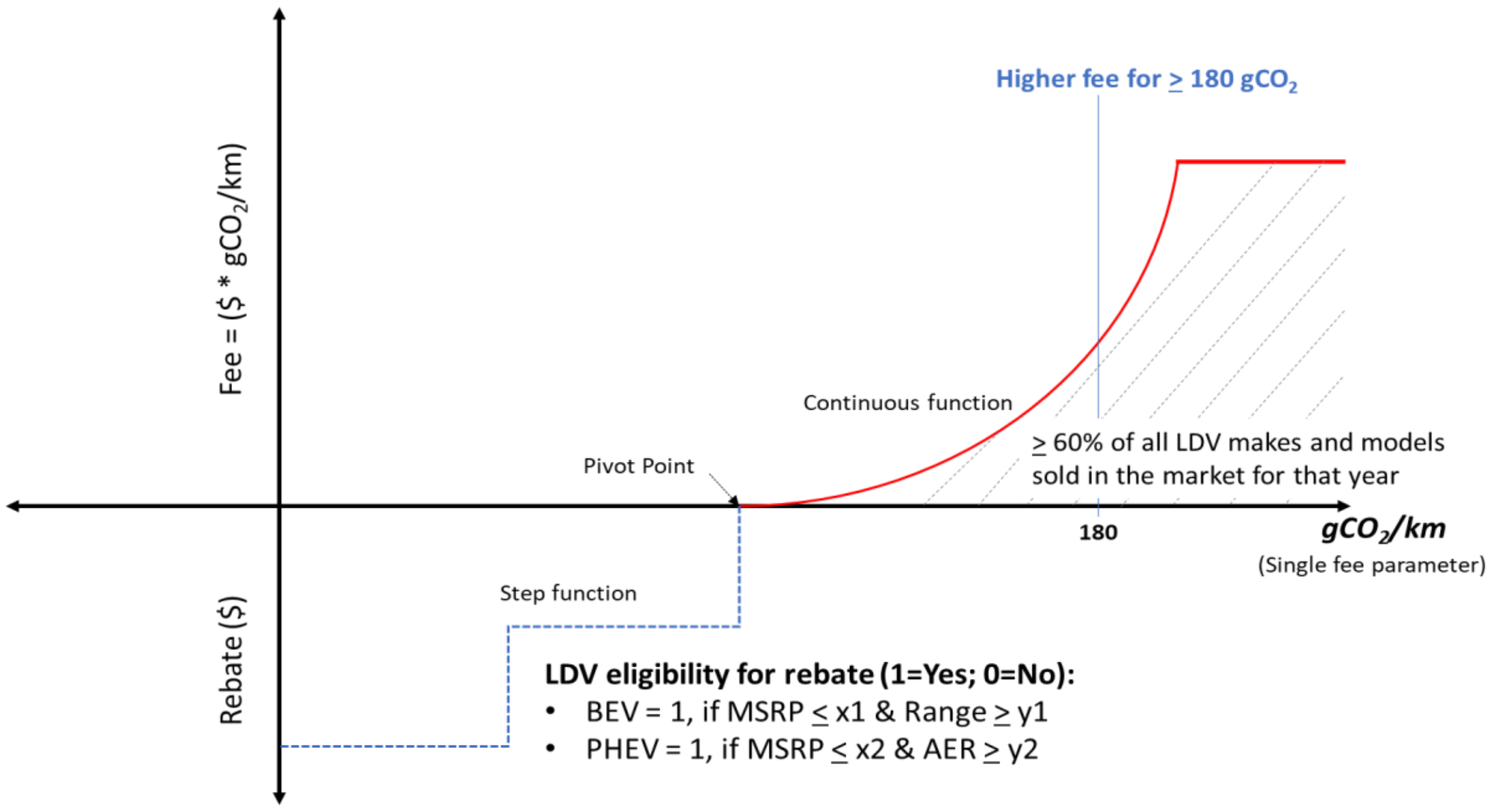

**Figure 28: Key elements of a feebate mechanism towards an inclusive EV transition**

## 7. Conclusions

Effectively, to achieve a ZEV transition, countries will have to re-align their feebate mechanisms in a manner that targets most of the ICE vehicles sold, while keeping in mind equity considerations for those in low-income groups and other disadvantaged communities. Having said that, reaching the goal of a ZEV transition will likely happen in phases, with a mix of PHEV and BEV sales as in present conditions, and then, a shift to only BEV and other ZEV technologies. As countries look to incentivize the transition to low emission vehicles, the financial burden of such a program can be limited, but with the growing urgency for a ZEV transition, the fiscal pressure for many countries can be significantly higher, given potentially long-term rebate requirements to sustain the transition and higher technology costs for alternate ZEV technologies. Feebates, while providing a mechanism to disincentivize vehicles (fees) with lower fuel efficiency and higher emissions, offers an opportunity to generate revenues from the fees to finance rebates for purchase of cleaner vehicles. The feebate mechanism is a good approach to raise the necessary capital for financing a ZEV transition, in combination with other regulatory mechanisms. They can also play a critical role in pushing manufacturers towards investing in ZEVs, thus, bridging the gap between TCO and price parity between EVs and ICEVs.

Based on the feebate design, it can be revenue neutral or revenue positive, the latter offering opportunities to utilize the additional funds, for example, to create basic public charging infrastructure, based on a charger to vehicles ratio, depending on an analysis of typical driving patterns in the region.

While the notion of an additional tax is often seen as a political risk, feebates are essentially redistributive in nature and adhere to the more fundamental principles of taxation, which is to create a public good to the best extent possible. In a feebate, the tax is on higher polluting vehicles, and if designed well, can offer rebates targeted towards assisting middle and low-income households to purchase electric vehicles. Today, the need for rebates for EVs is driven by the higher upfront purchase price even though the operating cost of an EV is almost half or lower compared to an ICEV.

Also, feebates need not be used in perpetuity. Once, price parity is achieved as the feebate pushes EV sales volumes to a critical mass, by impacting both consumer choices and manufacturer strategies, the rebate burden will decline significantly, and will be largely required for a smaller share of low-income households.

In the future, innovations in feebate design can be adopted to meet specific transportation goals. For example, the mechanism of estimating the fee itself can be changed, to include a $CO_2$ price and the average lifetime use of the vehicle. Feebates could also be adapted to promote shared vehicles and efficiency in passenger-miles, rather than vehicle-miles (as the ultimate purpose is to move people, not vehicles; in other words, to provide the best mobility to the most people with the fewest vehicles). In principle, feebates could account for the capacity utilization of vehicles, as large vehicles with many passengers may be more environmentally friendly than smaller vehicles with lower capacity. At the same time, VMT-based feebates could have equity considerations as lower income households tend to travel longer distances out of compulsion and not voluntarily. Other innovations could include feebate designs to be net-revenue generating and using a portion of those revenues towards maintaining road infrastructure or subsidizing public transit and other active modes.

Future analysis could include: (i) an econometric evaluation to estimate the effect of feebates on EV sales compared to other vehicle parameters and market conditions; (ii) evaluating different feebate designs within geographies while moving to a ZEV transition; and, (iii) evaluating a potential for a feebate design for other leading automotive markets such as the US or India, to facilitate an EV transition. Another aspect that can be considered for future analyses would be to evaluate changes in EV prices over time, as manufacturers meet regulations and realize scale and technology maturity benefits, which can play a significant role in consumer demand, in addition to vehicle taxes and purchase incentives (fees and rebates).

**Acknowledgements**

We would like to thank Prof. Daniel Sperling, Founding Director, Institute of Transportation Studies (ITS), UC Davis, for his guidance and inspiration for this analytical paper. We would also like to thank Prof. David L. Greene, University of Tennessee, Knoxville, for valuable insights through the course of developing this paper.

**Funding:** This work was supported by the National Center for Sustainable Transportation (NCST), a US Department of Transportation center hosted at the Institute of Transportation Studies, University of California Davis.

**CRediT author statement**

**Aditya Ramji:** Conceptualization, Data curation, Formal analysis, Investigation, Methodology, Project administration. **Lewis Fulton:** Conceptualization, Project administration, Supervision, Validation. **Daniel Sperling:** Conceptualization, Supervision, Validation.